\documentclass[12pt]{article}
\title{Nonperturbative  equation of state of quark-gluon plasma. }
\author{ Yu.A.Simonov\\
State Research
Center\\Institute of Theoretical and Experimental Physics, \\
Moscow, 117218 Russia}

\newcommand{\beq}{\begin{eqnarray}}
 \newcommand{\eeq}{\end{eqnarray}}
\newcommand{\be}{\begin{equation}}
 \newcommand{\ee}{\end{equation}}

 \def\la{\mathrel{\mathpalette\fun <}}
\def\ga{\mathrel{\mathpalette\fun >}}
\def\fun#1#2{\lower3.6pt\vbox{\baselineskip0pt\lineskip.9pt
\ialign{$\mathsurround=0pt#1\hfil ##\hfil$\crcr#2\crcr\sim\crcr}}}

\newcommand{{\SD}}{\rm SD}

\newcommand{\vex}{\mbox{\boldmath${\rm x}$}}
\newcommand{\vey}{\mbox{\boldmath${\rm y}$}}
\newcommand{\ver}{\mbox{\boldmath${\rm r}$}}
\newcommand{\vesig}{\mbox{\boldmath${\rm \sigma}$}}

\newcommand{\vep}{\mbox{\boldmath${\rm p}$}}

\newcommand{\vez}{\mbox{\boldmath${\rm z}$}}
\newcommand{\veS}{\mbox{\boldmath${\rm S}$}}
\newcommand{\veL}{\mbox{\boldmath${\rm L}$}}
\newcommand{\veR}{\mbox{\boldmath${\rm R}$}}

\newcommand{\vew}{\mbox{\boldmath${\rm w}$}}
\newcommand{\verho}{\mbox{\boldmath${\rm \rho}$}}

\newcommand{\lan}{\langle}
\newcommand{\ran}{\rangle}

\begin{document}
\maketitle
\begin{abstract}

The  paper is devoted  to the systematic derivation of
nonperturbative  thermodynamics of quark-gluon plasma, on the
basis of the background perturbation theory. Vacuum background
fields enter only in the form of field correlators, which are
known from lattice and analytic calculations. In the lowest order
in $\alpha_s$  the  purely nonperturbative $sQGP$ thermodynamics
is expressed through single quark and gluon lines (single line
approximation) which are interacting nonperturbatively  with
vacuum fields and  with other lines. Nonperturbative EOS is
obtained in terms  (of the absolute value) of fundamental
(adjoint) Polyakov loop $L_{fund(adj)}$ and spatial string tension
$\sigma_s(T)$. In the lowest approximation  the pressure for
quarks (gluons) has a simple factorized form $P_{q(g)}= P_{SB}
L_{fund(adj)} ,$ where  $ L_i$ describe the action of vacuum
colorelectric fields
 on particle trajectory.

\end{abstract}

\section{Introduction}

The study of the deconfined phase of QCD has a long history and
was based mainly on the idea  of  weakly interacting gas of quarks
and gluons, where perturbative and some nonperturbative  methods can be used, (see a  book in
\cite{1} and reviews in \cite{2}).

Recently, however, it was realized that there is a strong
interaction in the QCD vacuum above the deconfinement temperature
$T_c$, as witnessed by the RHIC experiments, (see \cite{3} for
reviews) and lattice calculations (see e.g. \cite{4}
 for a rewiew and further references).

 Independently of these developments, another approach was
 suggested in QCD almost two decades ago (\cite{5}, see \cite{6,7}
 for  a reviews), where the Nonperturbative (NP) dynamics, instead
 of the perturbative one, was at the basis of QCD, and systematic
 formalism was developed both for $T=0$ (\cite{8,9}) and for
 $T>0$ (\cite{10}-\cite{12}). According to this approach, and in
 agreement with existing lattice data, the new picture of the
 deconfined phase of QCD was suggested, where NP dynamics plays
 the dominant role. The latter is described by simplest (Gaussian)
 Field Correlators (FC) of color electric and color magnetic
 fields, $D^{E}(x), D^{H} (x), D_1^{E}(x) $ and
 $D_1^{H}(x)$ which can be found from lattice measurements
 \cite{13,14} or from analytic calculations \cite{15}. Here
 correlators $D^{E}(x), D^{H} (x)$ are of pure NP origin,
 while $D_1^{E,H}$ contain all perturbative diagrams \cite{16}.
 The phase transition was
 predicted to be due to the disappearance of  $D^{E} (x)$ ,
 which ensures colorelectric confinement, and  the transition
 temperature $T_c$  was calculated (in  terms of gluonic
 condensate) in good agreement with lattice data \cite{17}. As  a
 result above $T_c$ the  NP dynamics was predicted with
 colormagnetic confinement due to $D^{H} (x)$  and this
 prediction was supported later by detailed lattice measurements
 \cite{14,18}.

 This NP dynamics of  (color)magnetic confinement plus the NP
 component of $D_1^{E}
 $ (which survives the phase transition) suggest a picture which
 is much different from
  the popular perturbative approach.
 In particular, the strong interaction in systems of quarks and
 gluons, leading to the formation of e.g. $Q\bar Q$
bound state was predicted 15 years ago \cite{19}, and the role of
the spatial string tension $\sigma_s (T)$ for $T\geq T_c$ was
stressed in \cite{10, 17}, and supported  by lattice data
\cite{13,14,20}.

Recently, as a new challenge to the perturbative picture of
quark-gluon plasma, the Debye masses have been computed through
$\sigma_s (T)$ \cite{21}, in good agreement  with lattice data
\cite{22}.

An important dynamical role in the formalism of field correlators
 is played by Polyakov  loops,  and  the latter have  been
computed in terms of correlators  and the property of Casimir
 scaling for them was theoretically established in \cite{23}.
  Recently the Casimir scaling for five lowest representations
SU(3) was accurately checked on the lattice \cite{24}, thus
 supporting the quantitative accuracy of
the method also for $T>T_c$. At this point it is clear that a
systematic NP formalism for all phenomena in $T>0$ QCD is needed,
which should be a basis for calculation of EOS, Polyakov loops,
susceptibilities etc,.
 In the present paper,  we undertake this task, giving a detailed derivation of
the expressions for generating functions, pressure and internal
energies of quarks and gluons, where perturbative contributions
are considered as a series of Background Perturbation Theory
(BPTh) and NP effects due to NP background can  be  taken fully in
each term of $O(\alpha_s^n)$.

The  present paper is concerned with the lowest order
$O(\alpha_s^0)$ of BPTh and gives the pressure of quarks and
gluons  in the form, where the leading contribution is due to
single quark and gluon line interacting with the NP background
(Single Line Approximation (SLA)), while line-line NP interaction
is a correction. In SLA the pressure can be written in a simple
factorized form Eqs. (66), (67), where NP factors are of the
absolute value Polyakov loops,  while   colormagnetic background
fields enter in the line-line correction.

 Our main region of interest is the low
temperature region, $T_c<T<2T_c$, where nonperturbative effects
are most important, while at  higher temperatures one expects
possible connections of this  formalism with the well developed
perturbation theory now known to $O(g^6)$ (see \cite{25} and refs.
therein).

Two limitations are present in this  paper: we do not consider
chiral effects,  and possible accumulation of quark and gluon
bound states near $T_c$

The plan of the paper is as follows. In section 2 the basics of
BPTh and path integrals is given and free quark-gluon gas result
is derived as an illustration. In section 3 NP interaction of
quarks and gluons with background is expressed in the  form of
Polyakov loops (for colorelectric fields). Polyakov loops are
studied in section 4 and EOS is summarized in section 5. Section 6
is devoted to line-line interaction. Results and discussion are
given in the concluding section.

\section{ Background Perturbation Theory at $T>0$}

\subsection{Basic equations}

We start with standard formulae of the background field formalism
\cite{8,9} generalized to the case of nonzero temperature in
\cite{10}-\cite{12}. We assume that the gluonic field $A_\mu$ can
be split into the background field $B_\mu$ and the (valence gluon)
quantum field $a_\mu$ \be A_\mu=B_\mu+a_\mu, \label{23a} \ee both
satisfying the periodic boundary conditions  (PBC) \be
B_\mu(z_4,z_i) =B_\mu(z_4+n\beta, z_i);~~a_\mu(z_4,z_i)
=a_\mu(z_4+n\beta, z_i), \label{24a} \ee where $n$ is an integer
and $\beta=1/T$. The partition function can be written as \be
Z(V,T) =\lan Z(B)\ran_B \label{25a} \ee with \be Z(B)=N\int D\phi
\exp \left(-\int^\beta_0 d\tau \int d^3x L_{tot} (x,\tau)\right
)\label{4b}
 \ee
and where $\phi$ denotes all set of fields $a_\mu,\Psi, \Psi^+$;
$L_{tot}$ is   defined in \cite{10}-\cite{12} and for the
convenience of the reader  in Appendix 1. $N$ is a normalization
constant. Furthermore, in Eq. (\ref{25a}) $\lan~\ran_B$ means
averaging over (nonperturbative) background fields $B_\mu$. The
precise form of this averaging is not needed for our purpose.

Integration over the ghost and gluon degrees of freedom in Eq.
(\ref{25a}) yields the same answer as for the case $T=0$
\cite{19}, but where now all fields are subject to the periodic
boundary conditions (\ref{24a}).Disregarding for simplicity quark
terms and source terms in $L_{tot}$ in (\ref{4b}), one obtains
$$
Z(B)=N'(\det W(B))_{reg}^{-1/2}[\det(-D_\mu(B)D_\mu(B+a))]_{a=
\frac{\delta}{\delta J}}
$$
\be \times \left \{1+\sum^\infty_{l=1} \frac{S_{int} \left(a=
\frac{\delta}{\delta J}\right)^l}{l!} \right\}\exp \left(
-\frac12JGJ\right)_{J_\mu=D_\mu(B)F_{\mu\nu}(B)},
 \label{26a}
\ee where the valence gluon Green's function $G\equiv W^{-1}$ is
$G_{\mu\nu} = (\tilde D^2_\lambda \cdot \hat 1 + 2ig \tilde
F_{\mu\nu})^{-1}$, and the tilde sign here and below refers to the
operators in the adjoint representation, e.g. $\tilde
F^{bc}_{\mu\nu} \equiv i F_{\mu\nu}^a f^{abc}$. We can consider
strong background fields, so that $gB_\mu$ is large (as compared
to $\Lambda^2_{QCD}$), while $\alpha_s= g^2/4\pi$ in that strong
background is small at all distances. Moreover, it was shown that
$\alpha_s$ is frozen at large distances\cite{9}. In this case Eq.
(\ref{26a}) is a perturbative sum in powers of $g^n$, arising from
the expansion in $(ga_\mu)^n$.

In what follows we shall discuss the Feynman  graphs for the
thermodynamic potential  $F(T,\mu)$, connected to $Z(B)$ via \be
F(T,\mu)=-T\ln \lan Z(B)\ran_B .\label{27a} \ee It is  clear that
(\ref{27a}) contains all possible interactions, perturbative
 and nonperturbative (NP) between quarks and gluons, and in
particular creation or dissociation of bound states. It is
impossible to take into account all possible subsystems and
interaction between them, and it is imperative to choose the
strategy of approximations for the quark-gluon plasma as a
deconfined state of quarks and gluons.

We assume that the whole  system  of $N_g$ gluons and $N_q,
N_{\bar q}$ quarks and antiquarks for $T>T_c$ stays
gauge-invariant, as it was for $T<T_c$, however in case of
deconfinement and neglecting in the first approximation all
perturbative  and   NP interaction, any white system  will have
the same energy depending only on the number and type of
consistuents. In this case  $Z$ factorizes into a product of
one-gluon or one-quark (antiquark) contributions  and  we
calculate the corresponding thermodynamic potential.

This first step is called the Single Line Approximation (SLA) and
in the next two subsections we calculate the known results for
free gluon and quark gas in our  path-integral formalism following
\cite{10}.

However already in  SLA there exists a strong interaction  of  a
gluon (or $q, \bar q)$ with the NP vacuum fields. It consists of
colorelectric (CE)
 and colormagnetic (CM) parts, as shown in section 3. The CE part
 in the deconfinement case creates the individual NP self-energy
 contribution for  every gluon or quark, and this is  a factor of
 the corresponding  Polyakov loop discussed in section 4.
Note, that the quark (gluon) Polyakov loop factor is computed from
the gauge invariant $q\bar q(gg)$ Wilson loop, which for the NP
$D_1^E$ interaction separates into individual quark (gluon)
contributions equal to the modulus of the corresponding Polyakov
loop. Therefore our Polyakov loop factors are expressed through
color singlet 9heavy-quark) free energy $F_s(\infty, T)$ and the
popular $Z(N_c)$ symmetry is irrelevant for them.

 With the CM part the situation is more subtle. Strong CM fields
 in the deconfined QCD vacuum introduced in \cite{10}, \cite{17}
 and  measured on lattice \cite{13,14} create bound states of white
 combinations  of quarks and gluons, and this is formally beyond
 SLA. To take  these bound states into account one needs to use
 source terms, introduced in Appendix 1, and do calculations as
 shown in Appendix 2.

 Therefore  the CM fields are taken into account as two-body correlations, which are discussed in section
5 and Appendix 2.

 The SLA  result for the pressure  with the CE  is given in section 6.
 The effects of possible $q\bar q, 3q, gg, 3g, $ etc bound states
 which might be important nearby $T\geq T_c$,  as it is discussed
 in \cite{23}, are deferred to future publications.

\subsection{The lowest order gluon contribution}

To lowest order in $ga_\mu$ (keeping all dependence on $gB_\mu$
explicit) we have \be Z_0=e^{-F_0(T)/T}=N'\lan \exp
(-F_0(B)/T)\ran_B, \label{28a} \ee where using Eq. (\ref{26a})
$F_0(B)$ can be written as
\begin{eqnarray}
\frac{1}{T} F_0^{gl}(B) &=&\frac12\ln \det G^{-1} -\ln \det (-
D^2(B))= \nonumber
\\
&=&Sp\left\{ -\frac12 \int^\infty_0 \xi (s) \frac{ds}{s}
e^{-sG^{-1}} + \int^\infty_0 \xi (s) \frac{ds}{s}
e^{sD^2(B)}\right\}. \label{29a}
\end{eqnarray}

In Eq. (\ref{29a}) $Sp$ implies summing over all variables
(Lorentz and color indices and coordinates) and $\xi(t)$ is a
regularization factor. Graphically, the first term on the r.h.s.
of Eq. (\ref{29a})
 is a gluon loop in the background field, while the second term is
a ghost loop\footnote{In what follows we shall call the
contribution of gluon (ghost) Green's functions $G(D^{-2}(B))$ the
gluon(ghost) line contribution to distinguish it from the loop
expansion in standard perturbative analysis.}.

Let us turn now to  the averaging procedure in Eq. (\ref{28a}).
With the notation $f=-F_0(B)/T$, we can exploit in Eq. (\ref{28a})
the cluster expansion \cite{26}
 \begin{eqnarray}
\lan \exp f\ran_B&&=\exp \left (\sum^\infty_{n=1} \lan \lan
f^n\ran\ran\frac{1}{n!}\right) \nonumber
\\
&&=\exp \{ \lan f\ran_B+\frac12[ \lan f^2\ran_B- \lan
f\ran^2_B]+O (f^3)\}. \label{30a}
\end{eqnarray}
Below in this paper we consider mostly the leading  term in  the
cluster expansion, namely $\lan f\ran_B$, leaving the quadratic
and higher terms for the section 7.

To get a closer look at $\lan \varphi\ran_B$ we first should
discuss the thermal propagators of the gluon and ghost in the
background field. We start with the thermal ghost propagator and
write the  Fock-Feynman-Schwinger Representation (FFSR) for it
\cite{10} \be (-\tilde D^2)^{-1}_{xy}= \lan x|\int^\infty_0
dse^{sD^2(B)}|y\ran= \int^\infty_0 ds(Dz)^w_{xy} e^{-K}\tilde
\Phi(x,y). \label{31a} \ee Here $\tilde \Phi$ is the parallel
transporter in the adjoint representation along the trajectory of
the ghost: \be \tilde \Phi(x,y) =P\exp (ig \int\tilde B_\mu(z)
dz_\mu) \label{32a} \ee and $(Dz)^w_{xy}$ is a path integration
with boundary conditions imbedded (denoted by the subscript
$(xy)$) and with all possible windings in the Euclidean temporal
direction (denoted by the superscript $w$). We can write it
explicitly as
\begin{eqnarray}
(Dz)^w_{xy}&&= \lim_{N\to \infty}\prod^N_{m=1}
\frac{d^4\zeta(m)}{(4\pi\varepsilon)^2} \nonumber
\\
&& \sum_{n=0,\pm,...} \frac{d^4p}{(2\pi)^4}\exp \left[
ip_\mu\left( \sum^N_{m=1} \zeta_\mu(m)-(x-y)_\mu-n\beta
\delta_{\mu 4}\right)\right]. \label{33a}
\end{eqnarray}
Here, $\zeta(k)=z(k)-z(k-1),~~ N\varepsilon=s$. We can readily
verify that in the free case, $ \hat B_\mu=0$, Eq. (\ref{31a})
reduces to the well-known form  of  the free propagator
\begin{eqnarray}
(-\partial^2)^{-1}_{xy}=&&\int^\infty_0 ds\exp\left
[-\sum^N_1\frac{\zeta^2(m)}{4\varepsilon}\right ] \prod_m
\overline{d\zeta(m)} \sum_n\frac{d^4p}{(2\pi)^4} \nonumber
\\
 &&\times \exp \left [ ip_\mu \left ( \sum \zeta_\mu(m) -(x-y)_\mu
 -n\beta\delta_{\mu4}\right)\right]
\nonumber
\\
 =&&\sum_n\int^\infty_0\exp \left
 [-p^2s-ip(x-z)-ip_4n\beta\right] ds \frac{d^4p}{(2\pi)^4}
 \label{34a}
 \end{eqnarray}
 with
 $$
 \overline{d\zeta(m)}\equiv \frac{d\zeta(m)}{(4\pi\varepsilon)^2}.
   $$
   Using the Poisson summation formula
   \be
   \frac{1}{2\pi}\sum_{n=0,\pm1, \pm2...} \exp (ip_4 n\beta) =
   \sum_{k=0,\pm1,...} \delta(p_4\beta-2\pi k)
   \label{35a}
   \ee
   we finally obtain the  standard  form
   \be
   (-\partial^2)^{-1}_{xy}=\sum_{k=0,\pm1,...}
   \int\frac{Td^3p}{(2\pi)^3}\frac{\exp[-ip_i(x-y)_i-i2\pi
   kT(x_4-y_4)]}{ p^2_i+(2\pi kT)^2}.
   \label{36a}
   \ee

   Note that, as expected, the propagators  (\ref{31a}) and (\ref{36a})
 correspond to a sum of ghost paths with all possible windings
 around the torus. The momentum integration in Eq. (\ref{33a}) asserts
 that the sum of all infinitesimal "walks" $\zeta(m)$ should be equal
 to the distance $(x-y)$ modulo $N$ windings in the
 compactified fourth coordinate.
 For the gluon propagator in the background gauge we
 obtain similarly to Eq. (\ref{31a})
 \be
 G_{xy}=\int^\infty_0 ds (Dz)^w_{xy} e^{-K}\tilde
 \Phi_F(x,y),
 \label{37a}
 \ee
 where
 \be
\tilde \Phi_F(x,y)= P_F P\exp \left( 2ig \int^t_0 \tilde
 F(z(\tau))d\tau\right) \exp \left( ig \int^x_y \tilde
 B_\mu dz_\mu\right).
 \label{38a}
 \ee
The operators $P_F P$ are used to order insertions
 of $\hat F$ on the trajectory of the gluon.

 Now we come back to the first term in Eq. (\ref{30a}),
 $\lan\varphi\ran_B$, which can be representated with
 the help of Eqs. (\ref{31a})  and (\ref{37a}) as
 \be
 \lan F_0^{gl}(B) \ran_B= -T\int\frac{ds}{s} \xi(s) d^4
 x(Dz)^w_{xx} e^{-K}\left [ \frac12 tr \lan \tilde
 \Phi_F(x,x)\ran_B-\lan tr \tilde \Phi (x,x)\ran_B\right],
 \label{39a}
 \ee
 where $tr$ implies summation over Lorentz and
 color indices.
One can easily show \cite{10} that Eq. (\ref{39a}) yields for
 $B_\mu=0$ the usual result for  the free gluon gas:
 \be
 F_0^{gl}(B=0)=-Tf(B=0)=-(N^2_c-1)
 V_3\frac{T^4\pi^2}{45}.
 \label{40a}
 \ee

 \subsection{The lowest order quark contribution}

 Integrating over the quark fields in Eq. (\ref{25a})
 leads to the following additional factor in Eq. (\ref{26a})
 \be
 \det(m_q+\hat D(B'+a))=[\det (m_q^2-\hat D^2(B'+a))]^{1/2},
 \label{41a}
\ee where $B'_\mu =B_\mu-\frac{i\mu_g}{g} \delta_{\mu4},
\mu_q\equiv \mu$ is the quark chemical potential, and  $m_q$ is
the current quark mass (pole mass when next terms in $\alpha_s$
are considered).
 In the lowest approximation, we may omit
$a_\mu$ in Eq. (\ref{41a}). As a result we get a contribution from
the quark fields to the thermodynamic potential
 \be
 \frac{1}{T} F^q_0(B')=\frac12\ln\det (m_q^2-\hat D^2(B'))=-\frac12
 Sp\int^\infty_0\xi(s)\frac{ds}{s} e^{-sm_q^2+s\hat D^2(B')},
 \label{42a}
 \ee
 where $Sp$ has the same meaning as in Eq. (\ref{29a}) and
 \begin{eqnarray}
 \hat D^2=(D_\mu\gamma_\mu)^2=&&D^2_\mu(B')-gF_{\mu\nu}\sigma_{\mu\nu}
 \equiv D^2-g\sigma F;
\nonumber
\\
&&\sigma_{\mu\nu} =\frac{1}{4i}
 (\gamma_\mu\gamma_\nu-\gamma_\nu\gamma_\mu).
 \label{43a}
 \end{eqnarray}
 Our aim now is to exploit the FSR to represent Eq. (\ref{42a}) in a form
 of the path integral, as was done for gluons in Eq. (\ref{31a}). The
 equivalent form for quarks must implement the antisymmetric boundary
 conditions pertinent to fermions. We find \cite{10}
 \be
 \frac{1}{T} F^q_0(B')=-\frac12 tr \int^\infty_0\xi(s) \frac{ds}{s}
 d^4x
 \overline{(Dz)}^w_{xx}e^{-K-sm^2}W_\sigma(C_n),
 \label{44a}
  \ee
  where $tr$ implies summation over spin and color indices,
$$
  W_\sigma(C_n)=P_FP_A\exp\left (ig \int_{C_n} B'_\mu dz_\mu\right)
  \exp g\int ^s_0 \left (\sigma_{\mu\nu} F_{\mu\nu}\right) d\tau,
$$
and
 $$
 \overline{(Dz)}^w_{xy}=\prod^N_{m=1}\frac{d^4\zeta(m)}{(4\pi
 \varepsilon)^2}\times
$$\be
\times \sum_{n=0,\pm1,\pm2,...}\!\!\!
 (-1)^n\frac{d^4p}{(2\pi)^4}
 \exp \left [ip\left(\sum^N_{m=1}
 \zeta(m)-(x-y)-n\beta
 \delta_{\mu 4}\right)\right].
 \label{45a}
 \ee

It can readily be checked that in the case $B_\mu=0,~ \mu\equiv 0$
 the well known expression for the free quark gas is recovered, i.e.
 for $m_q=0$
 \be
 F^q_0({\rm free~quark}) =-\frac{7\pi^2}{180} N_cV_3T^4\cdot n_f,
 \label{46a}
 \ee
 where $n_f$ is the number of flavors. The derivation of Eq. (\ref{46a})
 starting from the path-integral form (\ref{44a}) is done similarly to
 the gluon case given in the Appendix of the last reference in Ref.
\cite{10}.

 The path  $C_n$ in Eq. (\ref{44a})  corresponds to $n$ windings in the
 fourth direction. Above the deconfinement transition temperature
 $T_c$ one sees in Eq. (\ref{44a}) the appearance of the  factor
 $$\Omega_{fund}=P \exp \left [ig \int^\beta_0 B_4(z)
 dz_4\right].$$ In the next section  it is shown that it will
 generate a gauge-invariant quantity
 \be
  L_{fund} =\frac{1}{N_c} \lan tr \Omega_{fund}\ran.
 \label{47a}
 \ee
 where $L_{fund}$ is the Polyakov loop in the fundamental
 representation.

For a nonzero $\mu$   and $n_f$ massless flavors with $B_\mu\equiv
0$ one has \be \lan F\ran=-\frac{4 N_cV_3 T^4}{\pi^2}{n_f}
 \sum^\infty_{n=1} \frac{(-1)^{n+1}}{n^4}cosh \left(\frac{\mu n}{T}\right).
 \label{48a}
 \ee

In what follows, we consider arbitrary fields $B_\mu(x)$ and
express $\lan F\ran$ through field correlators.

\section{ Nonperturbative dynamics of single quark and   gluon lines}

As was shown  in the previous section, Eq. (18), the pressure for
gluons can be written as (the $T$-independent term with $n=0$ is
subtracted, and the relation $P_{gl}V_3 =- \lan F_0 (B)\ran_B$ is
used), \be
 P_{gl} = {(N^2_c-1)} \int^\infty_0
 \frac{ds}{s} \sum_{n\neq 0} G^{(n)} (s)\label{P1}\ee
 and for quark one can write using (\ref{44a})
 \be
  P_q={2N_c}  \int^\infty_0 \frac{ds}{s} e^{-m_q^2s}
  \sum^\infty_{n=1}(-1)^{n+1} [S^{(n)} (s)+ S^{(-n)} (s)
   ].\label{P2}\ee

  Here $G^{(n)}$ and $S^{(n)}$ are the  proper-time kernels of  gluon and quark   Green's
  functions at nonzero $T$, which are defined as follows
  \be
  G^{(n)}(s) = \int (Dz)^w_{on} e^{-K}\hat{ tr}_a\lan W_\Sigma
  (C_n)\ran\label{P3}\ee

\be
  S^{(n)}(s) = \int \overline{(Dz)}^w_{on} e^{-K}\hat{tr}_f\lan W_\sigma
  (C_n)\ran\label{P4}\ee
where

\be \hat tr_f W_\sigma (C_n) =\frac{1}{N_c} tr_f W_\sigma
(C_n)\label{P5}\ee and
  \be
\hat{tr}_a  \lan W_\Sigma (C_n) \ran =\frac{tr_a}{(N^2_c-1)}
(\frac12 \tilde \Phi_F (C_n) -\tilde \Phi(C_n))\label{P6}\ee

In what follows we shall neglect  the first  exponent in (17) both
in $W_\Sigma$ and in $W_\sigma$, since this gives spin-dependent
interaction, which can be treated as a correction (we relegate
this treatment to the  Appendix 2). Therefore the calculation of
$G^{(n)}(s)$ and $S^{(n)}(s)$ needs first the evaluation of the
Wilson loop $\lan W(C_n)\ran$,

Our purpose is to calculate $G^{(n)}(s), S^{(n)} (n)$ using for
$\lan W(C_n)\ran $ the field correlators.

 To proceed one should look more carefully into the topology of
 the Wilson loop $W_{\Sigma,\sigma}(C_n)$, which is to be a closed
 loop in $4d$. At this point one should again emphasize that the
 basic states $|k>$ with the quark-gluon numbers $N_g^{k}, N^{(k)}_q, N_{\bar
 q}^{(k)}$ entering into the partition  function $Z$ are gauge
 invariant, $ Z=\sum_k\lan k| e^{-H/T}|k\ran$ and one  write for
 $|k\ran$ a generic decomposition of the type, $|k\ran =|(gg)
 (gg) (ggg) (q\bar q)...\ran$ where particles in parentheses form
 white combinations.

 Correspondingly (and neglecting interaction between white
 subsystems) for each white subsystem one has contribution
 proportional to the product, e.g. for the $(gg)$ or $(q\bar q)$
 system
 $$Z_{(12)} =\int \int d \Gamma_1 d \Gamma_2 \lan tr W_{\Sigma,
 \sigma} (C^{(1)}_n, C^{(2)}_n)\ran,$$ with $$ \int d\Gamma_i=\int
 ds_i e^{-K_i}(Dz^{(i)}).$$

 Here $\lan tr W_{\Sigma, \sigma} (C^{(1)}_n C_n^{(2)})\ran$ is
 the closed Wilson loop made of paths $(C^{(1)}_n C_n^{(2)})$ of
 two gluons (or $q\bar q$) and parallel transporters (Schwinger
 lines) in the initial and final states, necessary to make them
 gauge invariant.

 The important result of this paper, shown  below, is that
 colorelectric correlator $D^E_1$ yields factorized contribution,
 i.e. $Z_{(12)} \to Z_1 Z_2$, where each of the factors $Z_i$
 contains only a part of the  common loop in the form of the
 Polyakov loop factor (with the singlet free energy in the exponent).

 The contribution of colormagnetic fields does not factorize,
 however, as shown in Appendix 2, this contribution can be
 considered as a correction therefore in this section it will
 be neglected, leaving the topic to section 5.

 As we shall see, the 4d
path integral will be decomposed into a product, $D^4z \to D^3z
Dz_4$, and the interaction kernel also factorizes in the Gaussian
approximation $\lan W(C_n)\ran =\lan W_1\ran \lan W_3\ran$, which
leads finally to the relation (42). To proceed we shall use as in
\cite{23} the Field Correlator Method, where for any closed loop
$C_n$ one can apply the nonabelian Stokes theorem  and Gaussian
approximation and write (see \cite{6,7} for discussion and refs)
$$
  \hat{tr}_f \lan W(C^{(1)}_n C_n^{(2)})\ran =\frac{tr_c}{N_c} \lan P\exp ig \int_{C_n} B_\mu
dz_\mu\ran=$$\be=\frac{tr_c }{N_c} \exp \left( -\frac{g^2}{2}
  \int_{S_n}\int_{S_n} d \sigma_{\mu\nu} (u) d
  \sigma_{\lambda\sigma} (v) \lan F_{\mu\nu} (u) F_{\lambda\sigma}
  (v)\ran \right).
  \label{P8}
   \ee
 Here one should specify the surface $S_n$ with the surface
 elements $d\sigma_{\mu\nu}(u)$ in the integral on the r.h.s. of
 (\ref{P8}). For this analysis it is important to consider
 different terms in the sum over $\mu,\nu$ in (\ref{P8})
 separately.
For the field correlator in (34) one can use the  decomposition
(second ref. in \cite{5}) which for nonzero $T$ should be written
separately for colorelectric field $E_i$ and colormagnetic field
$H_i$ \cite{10,17}. With the definition  $H_k\equiv \frac12
\varepsilon_{ijk} F_{ij}$ one has

\beq \lefteqn{ {g^2 } \langle  \hat tr_f [ E_i (x) \Phi(x,y) E_k
(y) \Phi(y,x) ]
\rangle } \nonumber \\
& & = \delta_{ik} \left[ {D}^E + {D}_1^E + u_4^2 {\partial {D}_1^E
\over \partial u_4^2} \right] + u_i u_k {\partial {D}_1^E \over
\partial \vec{u}^2} ~, \nonumber \eeq  \beq \lefteqn{ {g^2 } \langle \hat tr_f [ H_i
(x) \Phi(x,y) H_k (y) \Phi(y,x) ]
\rangle } \nonumber \\
& & = \delta_{ik} \left[ {D}^H + {D}_1^H + \vec{u}^2 {\partial
{D}_1^H \over \partial \vec{u}^2} \right] - u_i u_k {\partial
{D}_1^H \over \partial \vec{u}^2} ~, \nonumber \eeq  \be {g^2 }
\langle \hat  tr_f [ E_i (x) \Phi(x,y) H_k (y) \Phi(y,x) ] \rangle
= -{1 \over 2} \varepsilon_{ikn} u_n {\partial {D}_1^{HE} \over
\partial u_4} ~. \label{EH} \ee

  We take first the term with $\nu=\sigma =4$ (note
 that by definition $\mu<\nu$ and $\lambda < \sigma$, see \cite{6,7}).
\be J_n^E \equiv \frac12 \int D_{i4, k4} (u,v) d\sigma_{i4} (u)
d\sigma_{k4}(v), \label{P9})\ee where we have defined $D_{\mu\nu,
\lambda\sigma} \equiv g^2 \lan \hat tr_f F_{\mu\nu} (u) \Phi(u,v)
F_{\lambda\sigma} (v) \Phi(v,u)\ran.$
 At this point one takes into
account that the surface $S_n$ is  inside the winding Wilson loop
for the gauge-invariant $q\bar q$ system, as it is done in
\cite{21} and Appendix 2. The contribution of $D_1$ leads to the
sum of terms for $q$ and $\bar q$ separately of the form
(\ref{P11}), which fact justifies the use of singlet $V_1(\infty)$
for single line quark (or gluon) term in $P_q$(or $P_{gl}$). For a
single quark one obtains in this way
$$
 J_n^E=\frac12 \int^{n\beta}_0 du_4 \int^{n\beta}_0 dv_4
 \int^\infty_0 \xi d\xi D_1^E (\sqrt{(u_4-v_4)^2+\xi^2})=$$
 \be
  =\frac12 n\beta \int^{n\beta}_0 d\nu
  \left(1-\frac{\nu}{n\beta}\right) \int^\infty_0
   \xi d\xi D_1^E (\sqrt{\nu^2+\xi^2}).\label{P10}\ee
Note that for $n=1$ one recovers the  expression for the Polyakov
loop obtained  in \cite{23}, namely \be
 L_{fund}=\exp (-\frac{1}{2T} V_1(\infty)), ~~ \frac{1}{2T} V_1 (\infty)
 = J^E_1\label{P11}
  \ee
   At this  point an important simplification
  occurs. Namely, originally the path  $C_n$ is a complicated path
  $z_\mu (\tau)$ in 4d, however the integral (37) over $d\xi$ is always from some point $z_\mu(\tau)$ on $C_n$ to
  infinity and
  does not depend on the exact form of $C_n$, and is the same as
  for the straight-line Polyakov loop. This is true, however, only
  for the $D_{i4.k4}$ and  not for $D_{ik,lm}$.

  Another important observation is that correlators $D_{\mu\nu,
  \lambda\sigma} (u,\nu)$ are not periodic function of
  $(u_4-v_4)$, in contrast to fields $F_{\mu\nu}(u)$ and
  $F_{\lambda\sigma} (v)$. This will be true also for path
  integrals from any point $x$ to arbitrary point $y$, (temperature Green's function)   and is a
  consequence of   the vacuum average of parallel transporter
  $\Phi(u,v)$ present in $D_{\mu\nu, \lambda\sigma}(u,v).$

  We turn now to other contributions in   (\ref{P8}),
  \be
  J^{EH}_n \equiv \frac12 \int D_{i4, kl} (u,v) d\sigma_{i4} (u)
  d\sigma_{kl}(v).\label{P12}
  \ee

This term  is treated in Appendix 2, and is shown to be $T$
independent, we shall neglect it in what follows.

We finally turn to the spatial term, which will play in what
follows a special role. Here the  main contribution comes from the
colormagnetic correlator  $D^H(u-v)$, which provides the area law
for the (closed) spatial projection  $A_3$ of the surface $S_n$.
Correspondingly we shall denote this term as (for $ A_3\gg
\lambda^2$, where $\lambda$ is the gluon correlation length,
$D^H(x)\sim \exp (-x/\lambda)$) \be \lan W_3(C_n)\ran = \exp
(-\sigma_sA_3)\label{P13}\ee where \be \sigma_s=\frac12 \int
D^H(x) d^2 x \label{P14}
 \ee
 and $A_3$ is the  minimal area of the spacial projection of the
 surface $S_n$. As in the $J^E_n$ case,
   one
   should start with white states of
    $q\bar q, gg$ or $3q$. As it is shown
     in Appendix 2, the colormagnetic vacuum $(D^H)$ acts only in the $L\neq 0$
  states of these systems, and the resulting contribution
   of $D^H$ does not separate into single-line terms,
    but rather acts pairwise (triple-wise for $3q$). Therefore one can account
     for color magnetic interaction in the higher (2-line or
     3-line) terms. One can  take it approximately averaging
     single line in the  gas of quarks and gluons, and denoting the
     corresponding term as Magnetic Loop Factor (MLF) $G_3(s)$ and
     $S_3(s)$ for gluons and quarks respectively. In what  follows
     we shall neglect colormagnetic interaction in the first
     approximation.
 As a result the $4d$ dynamics in (\ref{P3}),
(\ref{P4}) separates into $3d$ and $1d$, and one can write

\be G^{(n)} (s) \equiv G_4^{(n)} (s) G_3 (s)= \int (Dz_4)^w_{on}
e^{-K_4-J_n^E} G_3(s) \label{P15}\ee and similarly $S^{(n)}(s)
=S_4^{(n)}(s) S_3(s),$ with the free quark (gluon) factors

 \be
 S^{(0)}_3 (s) = G_3^{(0)}(s) = \int (D^3z)_{00} e^{-K_3}=
 \frac{1}{(4\pi s)^{3/2}}
 .\label{P16}
 \ee

One should notice at this point, that $G^{(n)}(s)$ and
$S^{(n)}(s)$, Eqs. (30) and (31) differ in spin-factors (17) and
(23), and in color group representation. Dropping spin-factors in
the lowest approximation, one obtains $S^{(n)}$ from $G^{(n)}$
replacing adjoint quantities (like $\sigma_s$(adj)) by fundamental
 ones ($\sigma_s(f))$. At this point one
  can use Casimir scaling \cite{23,24} to
  write $\sigma_s(adj) =\frac94 \sigma_s(fund)$.
 To compute $G_4^{(n)} (s)$ one can notice, that $J_n^E$ does not
 depend on $z_4$ and can be taken out of the integral, while the
 integral over $z_4$ can be taken exactly, namely splitting the
 proper-time interval $N\varepsilon =s, N\to \infty$ one can write

 $$
\int (Dz_4)^W_{on} e^{-K_4}=\prod^N_{k=1} \left( \frac{d\Delta
z_4(k)}{\sqrt{4\pi\varepsilon}}e^{-\frac{(\Delta z_4
(k))^2}{4\varepsilon} } e^{ip_4 \Delta z_4 (k)}\right)
e^{-ip_4n\beta} \frac{dp_4}{2\pi}=$$ \be
 = \int \frac{dp_4}{2\pi} e^{_ip_4 n\beta -p^2_4 s} =
\frac{1}{2 \sqrt{\pi s}} e^{-\frac{n^2 \beta^2}{4s}}\label{P17}
 \ee
 As a  result  one has for $G^{(n)}{(s)}$
 \be G^{(n)}(s) = \frac{1}{2 \sqrt{\pi s}} e^{-\frac{n^2\beta^2}{4s}
 -\hat J^E_n} G_3^{(0)}(s)\label{P18}\ee
 $$ S^{(n)} (s) =\frac{1}{2 \sqrt{\pi s}} ,
 e^{-\frac{n^2}{4sT^2}- J^E_n}~ S_3^{(0)}(s);~~ \hat J^E_n
 =\frac94 J^E_n$$
 In the next section we consider the Polyakov loop factor $J^E_n$

\section{Properties of Polyakov loops}

As it was discussed in the previous  section, Eq.(\ref{P1}), the
lowest order contribution to the free energy of the quark and
gluon contains the  path integral $S^{(n)}(s) $ and $G^{(n)}(s),$
Eqs. (\ref{P3}) and (\ref{P4}), which is expressed through the
Polyakov loop with $n$  windings, namely (cf Eq. (\ref{P18})). \be
 G^{(n)} (s) =\frac{1}{2 \sqrt{\pi s}} e^{-\frac{n^2\beta^2}{4s}}
 G_3^{(0)}(s)   L^{(n)}_{adj}
  \label{2'.1}\ee
  where
  \be
    L^{(n)}_{adj} \equiv  \exp  (-\tilde J^E_n),~~ \tilde J^E_n
   =\frac94  J^E_n
   \label{2'.2}\ee
   and  $J^E_n$ is given in (\ref{P10}),
   \be
   J^E_n =\frac{n\beta}{2} \int^{n\beta}_{0} d\nu
   \left(1-\frac{\nu}{n\beta} \right) \int^\infty_0 \xi d\xi
   D_1^E(\sqrt{\nu^2+\xi ^2}).\label{2'.3}
    \ee

    In what follows we shall discuss the following properties of
    Polyakov loops:
\begin{description}
    \item[a)] Dependence  of  $\ L^{(n)}$ on $n$. (in what
    follows $ L$ stands for $ L_i, i =$ fund,
    adj.)
    \item[b)] Renormalization of $ L^{(n)}$ .
    \item[c)] Calculation of  $L_{adj} $ and $L_{fund}$ through the field correlators.

\end{description}

{\bf a)} We start with $n=1$ and write as  in (\ref{P11}) \be
L_{fund}=
 \exp\left( -\frac{1}{2T} V_1 (\infty)\right),~~\frac{1}{2T} V_1
 (\infty)\equiv J_1^E\label{2'.4}\ee
where $V_1(r)$ was introduced in \cite{23} as a potential between
heavy quark $Q$ and antiquark $\bar Q$ at a distance $r$, obtained
from the correlator of the Polyakov loops. As it is seen in
(\ref{2'.3}) $V_1(\infty)$ (as well as $V_1(r))$ is calculated
directly through the correlator $D_1^E$ (for $T\geq T_c)$. The
latter has perturbative and nonperturbative contributions which we
 write as (we neglect posssible interference terms, which contribute near $r=0$
 \cite{27})
  \be D_1^E (x) = D^E_{1 pert} (x) +
D^E_{1nonp.}(x)\label{2'.5}
 \ee
 and it is clear that $D^E_{1pert}$ contributes to the color-Coulomb law  between $Q$ and $\bar Q$ and to the perturbative
 self-energy correction, which should be properly subtracted and
 renormalized, as discussed below in the  point b). The NP
 ingredient to  $ L$ is given by  $D_{1 nonp}^E$ and
 this is responsible for  the NP thermodynamics of
 the $qg$ plasma, which is the leading contribution considered in
 this paper.

 The form  of the $D_{1 nonp}^E $ has been found
 previously at $T=0$   analytically \cite{15,23,18}
 and on the lattice \cite{13,14} and it was shown  that $D_{1
 nonp}^E$ can be connected to the gluelump Green's function, the
 gluelump mass $M_1$, found in \cite{28}, defining the range of $D_{1 nonp}^ E$,
\be D_{1 nonp}^ E(x) \approx  const\exp (-M_1|x|), |x|\ga 1/M_1
\label{2'.6}\ee

In \cite{13,14} $D_1^E$ and the value of $M_1$ was computed on the
lattice, and in \cite{23} this calculation was compared to the
lattice data  \cite{29} for Polyakov loop correlator  showing a reasonable
agreement. Therefore one can use as an acceptable the value of
$M_1\approx 0.7 \div 1 $ GeV, which is comparable to the inverse
of gluon correlation length  $\lambda$ at  $T=0$;
$1/\lambda\approx 1 $ GeV.

The asymptotic behaviour (51) is also kept at $T>T_c$, as shown by
lattice data \cite{13,14}.

At this point one should use  the  behaviour (\ref{2'.6}) in
(\ref{2'.3}). It is important to stress, that $D_1^E$ is not
periodic in $\nu\equiv x_4-y_4$, we remind that \be D_1^E \sim
\lan tr E_i (x_4, \vex)\tilde  \Phi(x,y) E_i (y_4,
\vey)\ran\label{2'.7}\ee in contrast to $E_i(x)$, which satisfies
PBC $E_i(x_4+ n\beta, \vex)= E_i(x_4, \vex)$, and this is a
consequence of the presence of the factor $\Phi(x,y)$ which is not
periodic in $x_4-y_4$. Hence $D_1^E(\nu,\xi)$ is decreasing for
large $|\nu|$  exponentially as in (\ref{2'.6}). One can define in
the integral (\ref{2'.3}) two possible regimes: 1) $\beta M_1 \gg
1,$ and  2) $\beta M_1\ll 1$.

In the first case one can always  neglect in (\ref{2'.3})  all
dependence on $n$ except for the first factor, writing \be J^E_n
\approx \frac{n\beta}{2} \int^\infty_0 d\nu \int^\infty_0 \xi d\xi
D_1^E (\sqrt{\xi^2+\nu^2}) \cong nJ^E_1.\label{2'.8}\ee

This results in a simple relation
 \be
 L^{(n)}_i \cong (  L_i )^n, ~~ i={\rm adj,~ fund}.\label{2'.9}\ee

 In the second case $D_1^E$ extends over the region of several
 periods $n$, and for such $n$, that $n\beta M_1\la 1$, $J_n^E$
 behaves quadratically in $n$. Since $D_1^E$ is positive, this
 means that $\lan L^n\ran$ decreases at large $T, T\gg M_1$,
 faster than is given by the law  (\ref{2'.9}). Hence the general
 conclusion is that for  $T\ll M_1$ the law (\ref{2'.9}) holds,
 while for $T\gg M_1$ (\ref{2'.9}) is replaced by a stronger
 dependence of the type $ L^{(n)} \approx (L)^{n^2}$
 for the first $n$ terms satisfying $n<n* = \frac{T}{M_1}$.

 {\bf b)} We now turn to the point b) and discuss renormalization of
 Polyakov loop, which was introduced on the lattice in \cite{29}
 and was considered analytically in
\cite{23}.

In \cite{23} it was found that the Polyakov loop expressed through
the correlator $D_1^E$, can be written as \be  L_{fund} =\exp
(-\frac{V_1^{(np)}(\infty) +
V_1^{(pert)}(\infty)}{2T})\label{2'.10}\ee where
$V_1^{(np)}(\infty)$ is expressed through the NP (and regular at
$x=0$ part $D_{1nonp}^ E(x)$), while $V_1^{(pert)}(\infty)$  is
expressed through $ D_{1 pert}^ E(x)$ (cf. (\ref{2'.5})) and for
the lattice-type minimal distance a in the lowest order in $g^2$
is equal to  \be V_1^{(pert)}(\infty) =\frac{2C_2\alpha_s}{\pi}
\left( \frac{1}{a} -T\ln a\right)\label{2'.11}\ee
 Neglecting the next order and possible
 perturbative-nonperturbative contributions, one can define the
 renormalized Polyakov loop as
 \be
  L^{ren}_{fund}  \equiv \exp \left(
 -\frac{V_1^{(np)}(\infty)}{2T}\right)\label{2'.12}\ee

 This definition can be compared to the renormalization of color
 singlet free energy $F_s(r)$  used in \cite{29}, where it was noticed that
 $F_s(r)$ does not depend on temperature (at
 least for $T<2T_c$) for distances $r\leq 0.15$ fm. Thus
 normalization of the data at short distances allows to have a
 smooth limit to the $T=0$ case.

 Since $V^{np} (r) \sim r^2$ \cite{23,5}  at small distances, $F_s(r)$  actually coincides there  with
 $V_1^{(pert)} (r)$, and the divergent part of $V_1^{pert)} (r)$
 is in $V_1^{(pert)}(\infty)$. Hence  one can deduce, that normalization
 to the small distance part of $V_s(r)$ at $T=0 $, with
 \be
 F_s^{ren}(r) \cong V_1(r) -V_1^{(pert)}
 (\infty)\equiv V_1^{(ren)} (r)\label{2'.13}\ee
 is a justifiable renormalization procedure equivalent to dropping
 the divergent part, $V_1^{(pert)}(\infty)$ in the definition of
 $V_1^{(ren)}$. In what follows  we shall define $L^{ren}$ as in (\ref{2'.12}).

 One might ask how at all the (unphysical) $\frac{1}{a}$
 divergence can occur in the expression for the free energy
 (pressure), where physical quark and gluon Green's function
 enter. Indeed, we have used the path-integral form (\ref{P3}),
 (\ref{P4}), which is exact and corresponds to the standard perturbation theory at small distances.
 Therefore  self-energy parts due to gluon exchange have a
 normal logarithmic divergence which is eliminated by a standard
 renormalization procedure, setting the renormalized self-energy
 equal to the pole mass for quarks or zero for gluons. Therefore
 the divergent part is effectively eliminated from the Green's
 function when general form as in (\ref{P1}) is used. Note, that
 the factorized form (\ref{P15}) already used the fact that
 $J_n^E$ is a constant, not depending on $z_4, \vez$, which
 implies that the self-energy divergence is eliminated and only
 nonperturbative contribution to $ L$ is kept. Therefore
 $J_1^E$ in (\ref{P18}) is  already $V_1^{(np)} (\infty)/2T$ and
 corresponds to the renormalized Polyakov loop (\ref{2'.12}).

  At this point one should be careful to distinguish between the (theoretical)
    definition of  the Polyakov loop
  (57) and lattice values $L^{lat}_i$, since the latter, as
  discussed in \cite{23}, contain all possible excitations of the heavy-light system, not present in the
  light quark or gluon loop. For more discussion
  see last section of this paper.

{ \bf c)} Here we discuss calculation of $L_{fund}$ and $L_{adj}$
 through $D$ and $D_1$ both below and above $T_c$ As found in
 \cite{23}, $L_{adj}$ is
 \be  L_{adj}  =\exp \left\{ - \left(\frac94 V_D (r^*, T) +
 \frac98 V_1^{np} (\infty, T)\right)/T\right\} \label{2'.14}\ee
 where $V_D$ is given in Appendix A of \cite{23}, $T\la T_c$

\be V_D(r,T) =2 \int^\beta_0 d\nu (1-\nu T) \int^r_0 (r-\xi) d\xi
D^E (\sqrt{\xi^2+\nu^2})\label{2'.15}\ee
 and $r^*$ is an average distance between the adjoint line (static
 adjoint charge) and a gluon  of vacuum, forming together a bound
 state (a gluelump). In a rough approximation one  can replace
 $\frac94 V_D(r^*,T)\approx M_{glp},$ with $M_{glp}$-the gluelump
 mass of the order of $1.1\div 1.4$ GeV \cite{28}. In the similar
 way one has for a quark Polyakov loop
 \be  L_{fund} = \exp \{-(V_D(r^*_1,T) +\frac12 V_1^{np}
 (\infty, T) ))/T\},~~L_{adj} = ( L_{fund} )^{9/4}\label{2'.a15}\ee
 where $r^*_1$ corresponds to the average size of the heavy-light
 meson, $r^*_1\sim 1.2\div 1.4$ fm, and $V_D(r^*,T)\approx
 \varepsilon_{HL} \approx  0.6 \div 0.7$ GeV.

 Therefore both $ L_{adj} $ and $ L_{fund} $  for  $n_f >0$ are
 nonzero also below $T_c$, however small in this region. Above
 $T_c$ one has a Casimir scaling relation, derived first in
 \cite{23} as a consequence of  the fact that  Gaussian correlators $D, D_1$
 proportional to the charge squared.

 Recently the Casimir scaling as in (61) was accurately checked
 and confirmed on the lattice \cite{24} for five lowest
 representations of SU(3). This fact  shows that also for Polyakov
 loops as well as for Wilson loops  \cite{30} the lowest
 (quadratic) field correlators $D, D_1$ are dominant with accuracy
 of few percent. Hence the QCD vacuum is Gaussian dominant at all
 temperatures, (unless a finetuned cancellation of higher correlators takes place \cite{30}).

\section{Nonperturbative Equation of State (Single Line
Approximation)}

We are now in position to summarize our derivations of the
nonperturbative free energies for quarks and gluons in the Single
Line Approximation (SLA)  as follows: for gluons from (28), (45),
$$ P_{gl}^{SLA} =\frac{(N^2_c-1)}{2\sqrt{\pi}} \int^\infty_0
\frac{ds}{s^{3/2}} G_3 ^{(0)}(s) \sum_{n\neq 0}
e^{-\frac{n^2}{4T^2 s}}
 L_{adj}^{(n)}=$$ \be = \frac{(N^2_c-1)}{16{\pi^2}}
\int^\infty_0 \frac{ds}{s^{3}}  \sum_{n\neq 0} e^{-\frac{n^2}{4T^2
s}}  e^{-\tilde J^E_n}\label{74a}\ee where  $\tilde J_n^E$ is
given in (47), (48), (53).

For quarks of $n_f$
 flavours from (29), (45),
$$
P^{SLA}_q = \frac{2N_c}{\sqrt{\pi}} n_f \int^\infty_0
\frac{ds}{s^{3/2}} e^{-m_q^2s} S_3^{(0)} (s) \sum^\infty_{n=1}
(-1)^{n+1} e^{-\frac{n^2}{4T^2s} } L^{(n)}_{fund}cosh\frac{\mu
n}{T}=$$ \be= n_f
 \frac{N_c}{4{\pi}^2} \int^\infty_0
\frac{ds}{s^{3} } e^{-m^2_qs}  \sum^\infty_{n=1} (-1)^{n+1}
e^{-\frac{n^2}{4T^2s} }e^{- J^E_n}cosh\frac{\mu
n}{T},\label{76}\ee  The Polyakov loop exponents, $J_n^E$, are
expressed through the function $D_1^E$ in (37) and can be either
computed  analytically, as in \cite{23,18}, or taken from the
lattice data (see e.g. \cite{33}).

As a first check we consider the limit  of free gluons and quarks,
i.e. $ J_n^E \to 1.$

Integrating over $ds$ in (\ref{74a}), (\ref{76}) one obtains   for
the free gas  with  $m_q=0$, $\mu=0$ \be P_{gl}^{SB} (\sigma\to 0)
=\frac{\pi^2(N_c^2-1) T^4}{45},~~ P_q^{SB} (\sigma\to 0)
=\frac{7\pi^2 N_c  T^4}{180} n_f,\label{77}\ee where we have used
the standard sum values \be \sum^\infty_{n=1} \frac{1}{n^4}
=\frac{\pi^4}{90},~~ \sum^\infty_{n=1} \frac{(-1)^{n+1}}{n^4}
=\frac{7\pi^4}{720}.\label{78}\ee Another simple approximation
consists of neglecting all terms in the sum over $n$, except the
term with $n=1$.

The accuracy of this approximation can be estimated from  the  sum
(\ref{78}), where all terms with $n>1$ contribute less than 8\%,
and  for nonzero $J_n^E$ (for $T<1.5 T_c,  L_{fund}$ is less than
0.7 for $n_f=2$ \cite{37}) higher terms in $n$ are suppressed even
more.

Therefore with accuracy better than 10\% (for $\frac{\mu}{T}<1)$
one can suggest the following expressions for $P_{gl}^{SLA},
P^{SLA}_q$ \be P^{SLA}_{gl} =\frac{2(N^2_c-1)}{\pi^2}  L_{adj} T^4
\label{79}\ee \be P^{SLA}_q =\frac{4N_cn_f}{\pi^2}
\sum^\infty_{n=1} \frac{(-1)^{n+1}}{n^4}
(L_{fund})^n\varphi_q^{(n)} (T) cosh \frac{\mu n}{T} \approx
\frac{4N_c}{\pi^2}n_f L_{fund} T^4 \varphi_q^{(1)}(T)cosh
\frac{\mu}{T}\label{80}\ee where we have defined \be
 \varphi_{q}^{(n)}(T) =\frac{n^4}{16T^4} \int^\infty_0
\frac{ds}{s^3}  e^{-m^2_q s} e^{-\frac{n^2}{4T^2 s}} =\frac{n^2
m^2_q}{2T^2} K_2\left( \frac{ m_qn}{T}\right).\label{82}\ee

A representation  similar to (\ref{79}) was suggested a decade ago
in \cite{12},   for gluons  where instead of $\frac{2}{\pi^2},$ as
in (\ref{79}), the Stefan-Boltzmann factor $\frac{\pi^2}{45}$ was
used. With the lattice data for $ L_{adj}$ known at that time, the
resulting agreement of $P_{gl}^{SLA}$ was quite satisfactory.

In the  next paper \cite{34}  we  exploit new accurate data for
the renormalized Polyakov loops, fundamental with $n_f=2$
\cite{33} and $n_f =0$ \cite{29} and recent data for the adjoint
loop \cite{24}, showing a reasonable agreement with (\ref{79}),
(\ref{80}).

Summing up the series (\ref{80}) for  $m_q=0$ one  easily obtains
another form for $P^{SLA}_q$

 \be P^{SLA}_q (m_q=0) =\frac{n_f
T^4}{\pi^2} \left\{ \int^\infty_0
\frac{z^3dz}{e^{z-(\frac{\mu}{T}-J^E_1)}+1}+\int^\infty_0
\frac{z^3dz}{e^{z+\frac{\mu}{T}+J^E_1}+1}\right\}\label{83}\ee and
and for  $m_q\neq 0$ (see \cite{34} for details) one has instead
\be P_q^{SLA} (m_q) =\frac{n_fT^4}{\pi^2} \int^\infty_0
\frac{z^4dz}{\sqrt{z^2+\nu^2}} \left\{ \frac{1}{
e^{\sqrt{z^2+\nu^2} - \left( \frac{\mu}{T} - J^E_1\right)}+1}+
\frac{1}{ e^{\sqrt{z^2+\nu^2} +\frac{\mu}{T} +
J^E_1}+1}\right\}\label{84}\ee where $\nu=\frac{m_q}{T}$.

Another useful quantity to compare with lattice data is the
internal energy  density $\varepsilon,$ \be\varepsilon = T^2
\frac{\partial}{\partial T} \left( \frac{P}{T}\right)_V =
\varepsilon_{gl} +\varepsilon_q\label{85}\ee and from (\ref{79}),
(\ref{80}) we obtain (for $\mu=0$)
 \be {\varepsilon^{SLA}_{gl}} =\frac{2}{\pi^2} T^2
(N_c^2-1) \frac{d}{dT} (T^3L_{adj} )\label{86}\ee \be
{\varepsilon^{SLA}_{q}} =\frac{4N_c}{\pi^2} n_f T^2 \frac{d}{dT}
(T^3 L_{fund} \varphi_{q}^{(1)} (T))\label{87}\ee and finally for
the ``nonideality" of the quark-gluon plasma one obtains $$
\frac{\varepsilon^{SLA}-3P^{SLA}}{T^4}=\frac{2(N^2_c-1)}{\pi^2}
T\frac{d}{dT} ( L_{adj} )+$$ \be+ \frac{4N_c}{\pi^2}
T\frac{d}{dT}( L_{fund } \varphi_{q}^{(1)}(T)).\label{88}\ee It is
interesting to study the limit of large $T$ for
$\frac{P^{SLA}_i}{T^4}$ given in Eqs. (\ref{79}), (\ref{80}), and
compare it to the Stefan-Boltzmann value $\frac{P_{SB}}{T^4}
(T=\infty)$. For that purpose one needs the large $T$ limit of $
L_i, \varphi_q, i=gl,q$.

The situation is more transparent with the large limit of
$\varphi_q^{(1)}(T)$.

From (81) one can conclude that  $\varphi_q^{(1)}(T)$ tends
asymptotically to 1 with the (small) correction proportional to
$\frac{\tilde m^2_q}{T^2}$. As to $L_i$, from (47), (48), (49) it
is clear that the problem with $ L_i$ boils down to the large $T$
behaviour of $V_1(\infty, T)/T$. In this paper we take $ L_i$ as
given by lattice \cite{27,29,33} and analytic calculations,
\cite{15, 18,24} from which $ L_i$ grows with $T$ and reaches  the
value of 1 around $T=2T_c$, and may become larger than 1 beyond
3$T_c$. This is connected to the negative value of the lattice
free energy of static quark in this range of temperature
\cite{37}. The actual limit of $ L_i$ at infinite $T$ is connected
to the fact, how the magnetic confinement, viz. $\sigma_s(T)$
enters in $ L_i$, and this will be discussed in one of  the next
papers. In this way one expects  that all corrections to $ L_i -1$
should be proportional to some power of
$\frac{\sqrt{\sigma_s(T)}}{T} \sim O\left(\frac{1}{\ln
T/\Lambda_\sigma}\right)$, since $\sigma_s(T)$ is the only
nonperturbative dimensionful quantity which defines the dynamics
of high temperature QCD -- colormagnetic interaction \cite{17,19}.

\section{Beyond Single Line Approximation}

In previous sections one-particle contributions were considered,
i.e. SLA was used, however it was stressed, that some interactions
like the colormagnetic one, acts in white ensembles like $q\bar q,
gg$ etc. These latter contributions  generated by the source terms
listed in Appendix 1 were neglected  except for contributions of
$D_1^E$, which acts in $q\bar q(gg)$ but produced effectively
single-line term $V_1(\infty)$. In addition $D^E_1$ as well as
$D^H, D^H_1$ produce $q\bar q, gg, 3q$ etc bound systems
\cite{23,18,31} which may affect strongly dynamics at $T\geq T_c$,
as will be discussed below.

In general one can  write the contribution of the  pair (triple)
interaction, starting from the  pair (triple) Green's
function,e.g. for the $q\bar q$  or  $gg$ system as

\be G^{(n)}_{12} = \int d\Gamma_1 d\Gamma_2 \lan tr W_{\Sigma,
\sigma} (C^{(1)}_n, C^{(2)}_n)\ran\label{6.1}\ee with $d\Gamma_i =
\int^\infty_0 ds_i (Dz)^w_{x_iy_i} e^{-K_i}.$

Here the closed Wilson loop is formed in the same way, as
discussed at the beginning of section 3. From the $(q\bar q)$ or
$(gg)$ Green's function one can proceed to the derivation of the
two-body Hamiltonian  $H_{q\bar q} $ or $H_{gg}$ in the same way,
as it was done for $T=0$ in \cite{41}, \be G_{12}
(x_1x_2~|~y_1y_2) = \lan x_1x_2|e^{-H_{12}/T}|y_1
y_2\ran\label{6.2}\ee

Our purpose in this section is to discuss the  properties of the
resulting   $H_{q\bar q}, H_{gg}$ for $T>T_c$, and possible
consequences for the quark-gluon thermodynamics. Writing
$H_i=K_i+V_i,~~ i=(q\bar q), (gg), (qg),...,$ one can consider two
distinct  dynamics, the Colorelectric (CE) one due $D_1^E$, and
colormagnetic (CM) due to $D^H, D_1^H$.

In the CE case the interaction can be written as (see  \cite{23,
18} for  more discussion) \be V_{q\bar q}^{(CE)} (r,T) \equiv v_1
(r,T) \equiv V_1(r,T) -V_1(\infty, T)\label{6.3}\ee and \cite{23}
\be V_1(r,T) =\int^{1/T}_0 (1-\nu T) d\nu \int^r_0 \xi d\xi D_1^E
(\sqrt{\nu^2+\xi^2})\label{6.4}\ee

In (\ref{6.3}) we have subtracted $V_1 (\infty, T)$, which is
already accounted for in SLA in the form of Polyakov loops, see
Eq. ({55}).

Bound states of $q\bar q, gg, $ and $gq$ systems have been studied
in the field correlator formalism in \cite{23,18} and on the
lattice (see \cite{29} and refs. in \cite{23}, \cite{18}).

Characteristic feature of $v_1 (r,T)$ is  that it is short-ranged
$(r_{eff} \sim 0.3$ fm) and  can support bound $S$-states in
($c\bar c)_1, (gg)_1 (cg)_3, (gg)_8$ of weak binding,
$|\varepsilon|\la 0.14  $ GeV for $T\la 1.5 T_c$.

On the lattice in addition to $S$-states of ($c\bar c)_1, (b\bar
b)_1$ and light  $(q\bar q)$ also lowest $P$ state is claimed to
exist (see \cite{36P} for  a review).

It is  clear that weakly bound states should fast  dissociate in
the dense $qg$ plasma and therefore hardly produce any significant
effect on the production rate, however can be important for the
kinetic coefficients.

We now turn to  the CM case. Here dynamics of $q\bar q$ system was
carefully studied in \cite{31} using the formalism of \cite{41}.
The resulting interaction has several interesting features.

\begin{enumerate}
    \item  Interaction of quark and gluon systems with the CM
    vacuum occurs only in the states with the orbital momentum
    $L>0$.
    \item  Centrifugal barrier is effectively killed for $r^2\gg
    \frac{L(L+1)}{\sigma}$ and no spin-independent interaction is
    found in this region.

    \item Long-range spin-orbit forces are predominant at large
    $r$, \\ $V_{so} (r) \cong -{\veL \veS \sigma_s}/{\mu^2_{eff}
    r}$, where $\mu^2_{eff}$ is the effective mass squared to be
    determined from the extremum of the relativistic Hamiltonian (we disregard in this section the large $r$ damping of $V_{so}
    (r)$ due to string inertia in $\mu$.
\end{enumerate}

The latter can be written as  (the  $q\bar q$ system with equal
masses is treated here for simplicity) \be H_{ll}
=\frac{p^2_r+m^2}{\mu} +\mu+V_{SI}^{(ll)} (r) + V_{so}
(r)\label{6.5}\ee where the sum of potentials can be effectively
rewritten as (see \cite {31} for exact expressions) for $J=L+S$
\be V_{CM} (r) =\frac{L(L+1)}{(\mu+\sigma_s r) r^2}
-\frac{\sigma_s L}{2\mu(\mu+\sigma r)r}.\label{6.6}\ee Note that
$V_{CM} (r)\equiv 0$ for $L=0$, as was noted before. For large $r$
the first term decreases as $O\left(\frac{1}{r^3}\right)$.

Numerical analysis shows that the lowest bound states appear in
the situation when $r\la \mu/\sigma$, when $V_{CM} (r) $ can be
rewritten as \be V_{CM} (r) \cong \frac{L(L+1)}{\mu r^2}
-\frac{\sigma_sL}{2\mu^2 r} = \frac{L(L+1)}{\mu r^2} -
\frac{\alpha_{eff}}{r}\label{6.7}\ee with $\alpha_{eff}
=\frac{\sigma_sL}{2\mu^2}$.

Note that $\mu$ is to be found from the minimum of the Hamiltonian
eigenvalues, as it is always done in the einbein Hamiltonian
formalism \cite{41}. Moreover $V_{CM} (r)$ and especially
$V_{so}(r)$ is not treated as a correction: in contrast to the
standard $\frac{1}{M}$ expansion, the spin-dependent interaction
in the Field-Correlator Method is obtained in the framework of the
Gaussian (quadratic) approximation, which is proved to be
accurate,  see \cite{30}, \cite{24} for more details, and
\cite{40} for the derivation of spin-dependent interaction.

For the fixed $\mu$, one obtains the bound state mass from
$H_{ll}$ as \be
 M_{n_r, L} (\mu) =\frac{m^2}{\mu} +\mu - \frac{\mu
 \alpha^2_{eff}}{2(n_r+L+1)^2}.\label{6.8}\ee
 The physical mass of CM bound $(q\bar q)$ is obtained after an
 extremum in $\mu$ is found for some $\mu=\mu_0$. As was found in
 \cite{31}, this  extremum exists for $m\ga 0.25$ GeV, i.e. for
 $(b\bar b), (c\bar c)$ and possibly ($s\bar s)$ systems with the
  binding energies 0.007, 0.19 and 90 MeV for $\sigma_s =0.2$
  GeV$^2$.

  There is no extremum for $m<0.2$ GeV, which means that
  relativistic system of light $q\bar q$ is not stable (no lower
  bound for the mass).  The situation is the same as for the
  famous $Z>137$ problem in QED, where relativistic treatment of
  bound states for $\alpha_{eff}>1$ leads to inconsistencies (see
  \cite{38} for reviews). Physically the problem should be treated
  taking into account pair creation, i.e. in the one-body
  (heavy-light) Dirac formalism, as in \cite{38}, considering also
  lower continuum of negative energy states.

  In the framework of the relativistic (einbein) Hamiltonians
  \cite{41} this  be studied using the  matrix
  Hamiltonian,suggested  in \cite{39a}.

Thus light quarks in the ($q\bar q)$ system or together with heavy
antiquarks are unstable in the CM vacuum with respect to multiple
light pair creation. A similar, and even more pronounced situation
should occur for gluonic systems, like $(gg)$ or a gluon circling
around a heavy quark. Here Hamiltonian is the same as in
(\ref{6.5}) with the replacement $\sigma_s\to \frac94 \sigma_s$
and quark spin by  the gluon spin, so that \be \alpha_{eff}
(gluon, ~\mu) = \frac92 \alpha_{eff} (quark,~\mu).\label{6.9}\ee

Again the gluon systems are unstable and should be  stabilized by
creation of multiple $(gg)$ pairs, which means that the CM vacuum
stimulates numerous $(q\bar q)$ and $(gg)$ pair  creation.

Numerical estimates of these processes and  spectra of produced
quarks  and gluons are of immediate interest for the ion-ion
collisions and shall be treated  elsewhere.

At this point one should consider the effect of quark-gluon medium
on  the existence of  CM bound pairs. It is clear, that the  size
of  the pairs is  (from the minimum of $V_{CM}$ and
$\alpha_{eff}\approx 1)$  $r_{eff} \sim
\frac{L^{3/2}}{\sqrt{\sigma_s}}\sim 0.5$ fm $L^{3/2}$, and in the
medium of density $n$ with  closest neighbor distance $r_0\sim
n^{-1/3}$, one expects strong screening for $r_{eff}>r_0$. Hence
for $T>1.1 T_c$ when the energy density exceeds 2 GeV/fm$^3$, $r_0
\sim 0.5$ fm  the  high-$L$ levels are already screened. This
might explain a fast drop of entropy $S_\infty (T)$ beyond $1.05
T_c$ found on the lattice \cite{40a}, as well as large value of
$S_\infty(T)$ near $T_c$, since large -$L$, large-$J$ bound states
might  bring about a large  entropy.

\section{ Summary and conclusions}

We have used BPTH and field correlators to find the NP
contribution to the quark-gluon EOS. As  a result the free energy
(pressure) is represented as a double sum in powers of
$\alpha_s^n(T)$ and in  the number of interacting single-particle
trajectories. In  the lowest order in $\alpha^n_s$ one obtains the
purely NP corrections to the free (Stefan-Boltzmann) result, which
contains as a leading term  the single particle trajectories
interacting only with the NP vacuum (background) fields -- Single
Line Approximation discussed in chapters 1-6.

In the next terms of this series also NP interaction between quark
(gluon) trajectories is taken into account in chapter 6.

The most interesting result of the investigation above is that in
the leading approximation the influence of the  NP background can
be represented by two factors multiplying the Stephan-Boltzmann
result: one is the  Polyakov loop for quark  and another for gluon
taking into account colorelectric vacuum fields.

Thus the main contribution to the pressure $P(T)$ and all other
thermodynamic quantities is given in our approach by the NP impact
on the free otherwise trajectories of quarks and gluons.
Perturbative interaction is included in the next orders of
$\alpha_s^n, n\geq 1$, and yields additional correlation between
quark and gluon  trajectories (line-line interaction), which
together with the NP  correlation leads to a  possible formation
of quark and gluon bound states  in the region $T_c\leq T\leq 1.2
T_c$ and should be considered as a next step, discussed in chapter
6.

As discussed in section 5 and Appendix 2, the CM interaction
produces a rather strong potential with the dominant spin-orbit
term $V_{so}\sim -\frac{L\sigma_s}{\mu^2 r}$,  which for quarks
with momentum $p\sim \mu$ and $\frac{L}{r} \sim p$ yields the
plasma interaction ratio $\Gamma_q =\frac{|V_{so}|}{E_{kin}} \sim
\frac{\sigma_s}{T^2}$ and $\Gamma_q=\frac92 \Gamma_q$. Both
$\Gamma_q$ and $\Gamma_g$ are large and presume a strong
interacting plasma in the temperature region, where screening due
to high pressure is not operating. At large $T$ the Polyakov and
magnetic loop factors tend to unity, and  the pressure tends to
the Stefan-Boltzmann limit, as it is in the standard perturbation
theory. Comparison of our results, Eqs. (\ref{74a}), (\ref{76})
and (\ref{79}), (\ref{80}) to the lattice data is given in detail
in the next paper \cite{34}, where it will be shown, that a
reasonable agreement is achived already in the situation when
line-line interaction is neglected, as in (\ref{79}), (\ref{80})
in the region at $T>1.2 T_c$, while in the narrow region  $T_c\leq
T\leq 1.2 T_c$ one possibly needs to take into account line-line
corrections. This result is in agreement with an earlier
investigation (see the first reference of \cite{12}), where also a
reasonable agreement was found for the gluonic part of the
spectrum.

At this point it is interesting to compare two distinct pictures
of the sQGP. In the first one the  main dynamics  of sQGP is due
to interaction between the  quarks and gluons, which  should be
taken into account to  high orders, as in \cite{25}, with possible
partial resummation of the pertubartive series, see e.g.
\cite{38a}. It is important, that vacuum fields do  not  enter
directly into the picture,  however can be taken into account by
adjusting to the dimensionally reduced theory at larger $T$.

From this prospective the theoretical method suggested in the
present and next papers of this series, is approaching sQGP from
the opposite direction as compared to the perturbation theory.
Indeed, the main contribution is obtained from  the vacuum (NP
background) and perturbative  corrections are  treated as small at
all temperatures.One important result of the  proposed method is
the calculation \cite{21} of the Debye mass $m_D$ which is
expressed through the spatial string tension $\sigma_s,
m_D=c_D\sqrt{\sigma_s}$ with calculable coefficient $c_D$. A good
agreement with  the  lattice data demonstrated in \cite{21} gives
a strong support to the idea, that the NP vacuum is after all the
main source of dynamics both in the confined and in the deconfined
phase. The main outcome is the
 picture  of almost independent
quasiparticles.

This is in agreement with the lattice measurements of the
correlation between strangeness and baryon number or electric
charge, demonstrating almost zero effect for $T>T_c$ \cite{39}.

The method suggested in the present paper  can be easily extended
to find the phase diagram in the ($\mu, T)$ plane, which is done
in the next paper of this series. N.O.Agasian  actively
participated in the first stage of this work, and the author is
indebted to him for suggestions and discussions. The author is
grateful for useful discussions to K.G.Boreskov,  S.N.Fedorov,
A.B.Kaidalov, O.V.Kancheli, B.O.Kerbikov and M.B.Voloshin. This
work is supported by the grants RFBR 06-02-17012,  grant for
scientific schools NSh-843.2006.2 and the State Contract
02.445.11.7424.

This work is done with financial support of the Russian Federal
Agency of Atomic Energy.

\vspace{2cm}

{\it \bf Appendix 1. Background Perturbation Theory for $T>0$}\\

 \setcounter{equation}{0} \def\theequation{A1.\arabic{equation}}

We start with standard formulas of the background field formalism
\cite{8,9} generalized to the case of nonzero temperature. We
assume that the gluonic field $A_\mu$
 can be split into the background field $B_{\mu}$  and the quantum field
$a_{\mu}$
\begin{equation}
A_{\mu}= B_{\mu}+a_{\mu},\label{132}
\end{equation}
both satisfying periodic boundary conditions
\begin{equation}
B_{\mu}(z_4, z_i) = B_{\mu}(z_4+n\beta, z_i), a_{\mu} (z_4, z_i) =
a_{\mu} (z_4+ n\beta, z_i),\label{133}
\end{equation}

where $n$ is an integer and $\beta= 1/T$. It will be convenient in
what  follows to use as in \cite{8} the background Lorenz gauge
\be D_\mu(B) a_\mu =0,~~ D_\mu^{ca} (B) \equiv
\partial_\mu\delta_{ca} -ig T^b_{ca} B^b_\mu,\label{A1.3}\ee
and to split the gauge transformation as follows (in fundamental
representation) $$A_\mu(x)= U^+(x) (A_\mu(x)+\frac{i}{g}
\partial_\mu) U(x),$$
\be B_\mu(x) =U^+(x) (B_\mu(x) +\frac{i}{g}\partial_\mu)
U(x)\label{A1.4}\ee $$ a_\mu(x) = U^+ (x) a_\mu(x) U(x)$$ One can
see, that the form of gauge condition (\ref{A1.3}) is invariant
under gauge transformations  (\ref{A1.4}).
 The
partition function can be written as
$$ Z(V,T,\mu) =<Z(B)>_B\;,$$
\begin{equation}
Z(B)=N\int D\phi \exp (-\int^{\beta}_0 d\tau \int d^3x
L(x,\tau))\label{134}
\end{equation}
 where $\phi$ denotes all set of fields $a_{\mu}, \Psi, \Psi^+,N$ is a
normalization  constant, and the sign $<>_B$  means some averaging
over (nonperturbative) background fields $B_{\mu}$, the exact form
of this averaging is not needed for our purposes. Furthermore, we
have
\be L_{ tot}(x,\tau)=\sum^{8}_{i=1} L_i+ L(j^{(n)}, a_\mu,
\Psi,\Psi^+),\label{A1.6}\ee
where
\begin{eqnarray}
\nonumber L_1=\frac{1}{4} (F^a_{\mu\nu}(B))^2;  L_2=\frac{1}{2}
a_{\mu}^a W_{\mu\nu}^{ab} a_{\nu}^b,
\\
L_3=\bar{\Theta}^a (D^2(B))_{ab}\Theta^b; L_4=-ig\bar{\Theta}^a
(D_{\mu}, a_{\mu})_{ab}\Theta^b
\\
\nonumber L_5=\frac{1}{2}\alpha (D_{\mu}(B)a_{\mu})^2;
L_6=L_{int} (a^3,a^4)
\\
\nonumber L_7=- a_{\nu} D_{\mu}(B) F_{\mu\nu}(B);~
L_8=\Psi^+(m+\mu\gamma_4+\hat{D}(B+a))\Psi\label{135}
\end{eqnarray}

Here $\bar{\Theta},\Theta$ are ghost fields, $\alpha$-
gauge--fixing constant, $L_6$ contains 3--gluon-- and 4--gluon
vertices, and we keep the most general background field $B_{\mu}$,
not satisfying classical equations, hence the
 the appearance of $L_7$.

An additional term in (\ref{A1.6}), $L(j^{(n)},...)$ serves as a
generating functional for creation of white multiquark or
multigluon systems, which are singled out at low temperatures. One
can write
$$
L(j^{(n)}, a_\mu, \Psi, \Psi^+) = \int dx_1 dx_2
\{j^{(gg)}_{\mu\nu} x, x_1, x_2) tr (a_\mu (x_1) \tilde \Phi(x_1,
x_2) a_\nu(x_2))+$$ \be+ j^{(q\bar q)}_\lambda (x, x_1, x_2) tr
(\Psi^+ (x_1) \gamma_\lambda\Phi (x_1, x_2)
\Psi(x_2))\}+...\label{A1.8}\ee

Here $\tilde \Phi(x_1,x_2)$ is the parallel transporter in adjoint
representation (marked by tilde) made of field $B_\mu$ only \be
\tilde \Phi (x_1, x_2) =P\exp (ig \int^{x_1}_{x_2} \tilde B_\mu
(z) d z_\mu)\label{A1.9}\ee $\Phi(x_1, x_2)$ is the same but in
the fundamental representaion. One can see that $L$ and $j^{(n)}$
are gauge invariant. Differentiating in $j^{(n)}$ one generates
white states of quark and gluons which enter in the total sum over
$n$ \be Z(V,T) =\sum_n\lan n|\exp(-\beta H) |n\ran
\label{A1.10}\ee

 The inverse gluon propagator in the background gauge is
\begin{equation}
W^{ab}_{\mu\nu} =- D^2(B)_{ab} \cdot \delta_{\mu\nu} - 2 g
F^c_{\mu\nu}(B) f^{acb}\label{136}
\end{equation}
where
\begin{equation}
(D_{\lambda})_{ca} = \partial_{\lambda} \delta_{ca} - ig T^b_{ca}
B^b_{\lambda} \equiv
\partial_{\lambda} \delta_{ca} - g f_{bca} B^b_{\lambda}\label{137}
\end{equation}

We consider first the case of pure  gluodinamics, $L_8\equiv 0$,
and omit the source term, $L$ in (\ref{A1.6}).

Integration over ghost and gluon degrees of freedom in (\ref{134})
yields
\begin{eqnarray}
\nonumber Z(B) =N'(\det W(B))^{-1/2}_{reg} [\det (-D_{\mu}(B)
D_{\mu}(B+a))]_{a=\frac{\delta}{\delta J}} \times
\\
\times \{ 1+ \sum^{\infty}_{l=1} \frac{S_{int}^l}{l!} (a=
\frac{\delta}{\delta J}) \} \exp (-\frac{1}{2} J
W^{-1}J)_{J_{\mu}= \;\;\;\;\;D_{\mu}(B)F_{\mu\nu}(B)}\label{138}
\end{eqnarray}

\vspace{2cm}

{\it \bf Appendix 2.~~Calculation of the Color Magnetic Interaction }\\

 \setcounter{equation}{0} \def\theequation{A2.\arabic{equation}}

Taking into account any interaction between gluons or quarks, and
in particular when the colormagnetic interaction  is considered,
one should recur to gauge invariant initial and final states
$\Psi$ in the Matsubara expansion for $P_g, P_q$, and define
$\Psi_{gg}, \Psi_{q\bar q}$ as follows

\be \Psi^{(0)}_{gg} (x,y) =tr (\hat a_\mu (x) \hat \Phi (x,y) \hat
a_\nu (y))\label{A3.1}\ee \be \Psi^{(n)}_{gg} (x',y') = \Psi_{gg}
(x+n\beta e_4, ~~ y+ n\beta e_4).\label{A3.2}\ee For the
double-line term in (93) one obtains \be \lan
Z(B)\ran_B^{two~line} = \int d\Gamma d\Gamma' \lan tr
W_{\Gamma\Gamma'}\ran \label{A3.3}\ee where $W_{\Gamma\Gamma'}$ is
a rectangular Wilson loop, formed by parallel transporters $\hat
\Phi(x,y), \hat \Phi (x', y')$ and gluon paths $(x,x')$ and
$(y,y')$.

First of all one can calculate $\lan tr W_{\Gamma\Gamma'}\ran$
using Gaussian approximation and cluster expansion \cite{5,6}. For
fundamental charges ($ q\bar q$ system) one has \be \lan
\frac{tr_c}{N_c} W_{\Gamma\Gamma'}\ran = \exp \left(
-\frac{g^2}{2} \int d\pi_{\mu\nu} (u) d\pi_{\lambda\sigma} (v)
\lan F_{\mu\nu} (u)  F_{\lambda\sigma}
(v)\ran\right)\label{A3.4}\ee where we have  omitted $\Phi(u,v),
\Phi(v,u)$ in field correlator for brevity and defined \be
d\pi_{\mu\nu} (u) = ds_{\mu\nu} (u) - i \sigma_{\mu\nu}
d\tau\label{A3.5}\ee and $ds_{\mu\nu} (u)$ is the surface element,
and $d\tau$ is proper time integration present in $W_\sigma$, Eq.
({23}).

We start with the terms coming from the product $ds_{\mu\nu} (u)
ds_{\lambda\sigma}(v)$ in (\ref{A3.4}), while all other terms have
been treated in \cite{40} and we shall write the resulting
equation at the end of this Appendix.

In what follows we shall derive as in \cite{41} the effective
Hamiltonian for the $q\bar q (gg) $ system, but taking into
account that colorelectric string tension $\sigma_E$ and the
colormagnetic one $\sigma_H \equiv \sigma_s$ can be different and
one of them $(\sigma_E)$ may vanish. In what follows we  are
closely following the paper \cite{31}.

With the notations of ({35}), ({36}) one can  write
$$
\lan \hat tr_f W_{\Gamma\Gamma'}^{(n)}\ran =\exp \left( -\frac12
\int \int  ds_{\mu\nu} (u) ds_{\lambda\sigma} (v) D_{\mu\nu,
\lambda\sigma} (u,v)\right)=$$ \be =\exp (-J_n (D^E_1) - J^{EH}_n-
J_n^H-J_n(D^E))\label{A3.6}\ee where $J_n(D_1^E)\equiv 2J^E_n$ and
$J_n^E$ is given in ({36}), while $J_n^{EH}$ is  $$J_n^{EH} \equiv
\frac12 \in\int D_{i4,kl} (u,v) d\sigma_{i4} (u) d\sigma_{kl}
(v)=$$ \be= \frac14 \int\int (\delta_{ik} w_l-\delta_{il}
w_k)\frac{\partial D_1^{EH}(w)}{\partial w_4} ds_{i4} (u)
ds_{kl}(v)\label{A3.7}\ee with $w_\mu= u_\mu-v_\mu$. The integral
$du_4$ in  (\ref{A3.7}) reduces to $D_1^{EH}(n\beta-v_4)
-D_1^{EH}(-v_4)$ and is small for $T\ll
\frac{n}{\lambda_{EH}}=nM_{EH}$, moreover this integral vanishes
for any flat surface.

We shall not discuss this term in what follows and now turn to the
term $J_n(D^E)$, which can be written as  $$J_n (D^E) =\frac12
\int \int ds_{i4} (u) ds_{k4}(v) D_{i4, k4}(u,v)=$$ \be= \frac12
\int\int du_4 dv_4 du_i dv_i D^E \left(\sqrt{(u_4-v_4)^2 +
(u_i-v_i)^2}\right)\label{A3.8}\ee

Introducing variables $\ver, \verho$
 \be \ver =\vex-\vey, ~~
\verho =[\ver\times \dot{ \vew}],~~\vew =\beta \vex+
(1-\beta)\vey,\label{A3.9}\ee where we have parametrized the
Wilson surface $W_{\Gamma\Gamma'}$ with $0\leq t \leq n/T;~~ 0\leq
\beta\leq 1$, so that the surface elements are \be ds_{i4} (u)
ds_{i4} (v) =\ver (t) \ver (t') dt dt' d\beta
d\beta'\label{A3.10}\ee while \be ds_{ik} (u) ds_{ik} (v) =2
\verho(t) \verho(t') dt dt' d\beta d\beta'.\label{A3.11}\ee One
can also write using induced metrics $g_{ab}$ \be (u-v)^2 =
(u(t,\beta) -v(t',\beta'))^2 = g^{ab} \xi_a \xi_b, ~~\xi_1 =
t-t',~~ \xi_2 =\beta-\beta'\label{A3.12}\ee and $\det
g=r^2+\rho^2$. In this way one finally obtains \be
J_n(D^E)=\sigma_E \int^{n/T}_0 dt \int^1_0 d\beta
\frac{r^2}{\sqrt{r^2+\rho^2}}\label{A3.13}\ee

\be J_n(D^H)=\sigma_H \int^{n/T}_0 dt \int^1_0 d\beta
\frac{\rho^2}{\sqrt{\rho^2+r^2}}.\label{A3.14}\ee In case when
$T=0$, one has $D^E=D^H,~~ \sigma_E=\sigma_H$ and one obtains

\be J_n(D^E) +J_n(D^H) =\sigma\int^{n/T}_0 dt \int^1_0 d\beta
\sqrt{r^2+\rho^2}\label{A3.15}\ee where the standard Nambu-Goto
action used in \cite{41} is  $\sqrt{\dot w^2 (w')^2-(\dot w_\mu
w'_\mu)^2}=\sqrt{r^2+\rho^2}$.

 In the action for equal mass quarks, $m_q=m_{\bar
q} =m,$  taking into account that \cite{41} \be K_q +K_{\bar q}
=\int^{T_0}_0 \frac{\mu}{2} (\dot{\vex^2}+\dot{\vey^2}) dt+
\int^{T_0}_0\mu dt\label{A3.16}\ee one finally obtains the
Hamiltonian for  $T=0$ \cite{41} \be H=\frac{p^2_r+m^2}{\mu} +\mu
+\frac{\veL^2/r^2}{2\left(\frac{\mu}{2} +\int\left(
\beta-\frac12\right)^2 \nu d\beta\right)} +\int \frac{\sigma^2
d\beta}{2\nu} r^2+\frac12 \int\nu d\beta\label{A3.17}\ee where the
einbein parameter $\nu$ is to be found from the extremum
condition.

For the case of electric deconfinement, when $\sigma_E=0$ and
$\sigma_H\neq 0$, one obtains instead \cite{31} \be
H=\frac{p^2_r+m^2}{\mu} +\mu +\int^1_0 d\beta \left(
\frac{\sigma^2_1r^2}{2\nu} +\frac{\nu}{2} +\sigma_2
r\right)+\label{A3.18}\ee
$$+ \frac{L(L+1)}{2r^2\left[ \frac{\mu}{2} + \int^1_0 d\beta
\left(\beta -\frac12\right)^2\nu\right]}$$ and in the case when
one mass is much larger than $\sqrt{\sigma_H},\sqrt{\sigma_E}$ one
obtains for the heavy-light system with both $\sigma_H, \sigma_E$
nonzero \cite{31}. \be H_{HL} = \frac{p^2_r+m^2}{2\mu}
+\frac{\mu}{2} +\int^1_0  d\beta \left( \frac{\sigma^2_1
r^2}{2\nu} +\frac{\nu}{2} +\sigma_2 r\right) +
\frac{L(L+1)}{2r^2[\mu +\int^1_0 d\beta \beta^2\nu]}\equiv H_{kin}
+h_{HL}\label{A3.19}\ee where we have  defined $H_{kin} \equiv
\frac{p^2_r+m^2}{2\mu} +\frac{\mu}{2}$, and \be \sigma_1 =\sigma_H
+ \eta^2(\sigma_H-\sigma_E),~~ \sigma_2=2\eta
(\sigma_E-\sigma_H).\label{A3.20}\ee

Here $\eta$ is another einbein parameter (cf \cite{41}).  Consider
first the case when $L=0$. Taking extremum of (\ref{A3.18}),
(\ref{A3.19}) with respect to $\nu$ one has \be H(L=0)= H_{kin} +
r\int^1_0 d\beta [\sigma_E +
(\sigma_H-\sigma_E)(1-\eta)^2]\label{A3.21}\ee
 and taking extremum in $\eta$, which yields $\eta_0=1$ one has
 for both Hamiltonians (\ref{A3.18}), (\ref{A3.19})
 \be
 H(L=0)=H_{kin} + \sigma_E r.\label{A3.22}\ee
 Hence for $\sigma_E=0, \sigma_H\neq 0$ and $L=0$ magnetic
 ``confining'' vacuum  does not  actually confine quarks (or
 gluons). Hence quotation marks for magnetic ``confinement''.

 Let us now turn to the case of $L\neq 0, \sigma_E=0$ and consider
 large $L$ limit, when one expects as in \cite{41} that $\nu_0\gg
 \mu_0$, where the subscript refers to the  extremum values of
 $\nu,\mu$. In this case taking extremum in $\nu$ in (\ref{A3.19})
 and neglecting $\mu$ in the last term on the r.h.s. of
 (\ref{A3.19}) one obtains
 \be
 \nu_0 =\left[\sigma_H^2(1+\eta^2)^2r^2 +
 \frac{3L(L+1)}{r^2}\right]^{1/2}
 \label{A3.23}\ee
 and for the $\eta_0$ - extremum value of $\eta$, one has an
 equation
 \be
 (\eta_0^4-1) (\eta^2_0+1)=\frac{3L(L+1)}{\sigma^2_Hr^4} \equiv
 c\label{A3.24}\ee
 with the solution
 $$\eta^2_0 =-\frac13 +\left\{ \frac12 \left(\frac{16}{27}
 +c\right)+\frac{\sqrt{c(c+32)}}{54}\right\}^{1/3}+$$
\be +\left\{ \frac12 \left(\frac{16}{27}
 +c\right)-\frac{\sqrt{c(c+32)}}{54}\right\}^{1/3}\label{A3.25}\ee

 The resulting Hamiltonian is
 \be
 h_{HL} =\sqrt{\sigma^2_H(1+\eta^2_0)^2 r^2 +\frac{3L(L+1)}{r^2}}
 - 2 \eta_0 \sigma_Hr \label{A3.26}\ee
 where $\eta_0$ is defined in (\ref{A3.25}).
 It is easy to check that for $L=c=0$ one  obtains $\eta_0=1$ and
 hence $h_{HL} (l=0)=0$, in  agreement with (\ref{A3.22}) for
 $\sigma_E=0$.

 For small $c$ (\ref{A3.26}) and (\ref{A3.25}) yield
 \be
 h_{HL} = \frac{3L(L+1)}{4\sigma_Hr^3},~~ 3L(L+1) \ll \sigma^2_H
 r^4.\label{A3.27}\ee

 In the opposite limit, $c\gg 1 (3L(L+1)\gg(\sigma_Hr^2))$ one
should take into account also $\mu$ in the denominator of $h_{HL}
$ in (\ref{A3.19}) and  one obtains the standard centrifugal term
 \be
 h_{HL} =\frac{L(L+1)}{2\mu r^2}.\label{A3.28}\ee
One can see, that the centrifugal barrier of light particle is
essentially destroyed by the magnetic vacuum for  $r>r_0$,
$r_0=\left(\frac{3L(L+1)}{\sigma^2_H}\right)^{1/4}$ while for
$r<r_0$ the barrier is essential .

The case of equal mass particles (light-light or heavy-heavy) is
done in the same way as above, the only difference being that
instead of $3L(L+1)$ one has $12 L(L+1)$ which should be
substitued in all Eqs. (\ref{A3.23}-\ref{A3.28}). Thus the
conclusion for the spin-independent part of the Hamiltonian, Eqs.
(\ref{A3.18}), (\ref{A3.19}), is that in magnetic vacuum the
centrifugal barrier is effectively destroyed for $r>r_0$, which
fact stresses  the importance of large distance spin-dependent
interaction -- to  be discussed now.

The  SDI is obtained from (\ref{A3.4}), when one considers
products of $ds_{\mu\nu}\cdot  \sigma_{\rho\lambda} d \tau$  ( for
spin-orbit forces) and $\sigma_{\mu\nu} d\tau
\sigma_{\rho\lambda}d\tau'$(for spin-spin and tensor forces), (see
\cite{40} and \cite{43} for details of derivation).

The resulting SDI can be represented in the Eichten-Feinberg from
\cite{44}\footnote{Here one should take into account that each
factor $ds_{ik}$  contains $\verho$ and  hence $\dot{\vew}$ and
each $\dot{z}_{i}$ for particle $i$ yields as in (\ref{A3.19}) the
factor $\hat \mu_i\equiv \mu_i+\int^1_0 \nu(\beta) \beta^2 d\beta$
 in  the denominator instead of $\mu_i$, while each $\vesig_i$ brings about a
 factor $\frac{1}{\mu_i}$. This additional $\nu(\beta)$ term in $\hat \mu_i$
 was hever taken into account before, and is  inserted in \cite{31} and below in (\ref{A3.40}), (\ref{A3.41}).}:

$$ V_{SD}^{(diag)}(R)=(\frac{\vec{\sigma}_1\vec L_1}{4\mu_1^2}-
\frac{\vec{\sigma}_2\vec
L_2}{4\mu_2^2})(\frac{1}{R}\frac{d\varepsilon}{dR}+\frac{2dV_1(R)}{RdR})+
$$
\be + \frac{\vec{\sigma}_2\vec L_1- \vec{\sigma}_1\vec
L_2}{2\mu_1\mu_2}\frac{1}{R}\frac{dV_2}{dR}+\frac{
\vec{\sigma}_1\vec {\sigma}_2V_4(R)}{12\mu_1\mu_2}+\frac{(3
\vec{\sigma}_1 \vec R\vec{\sigma}_2 \vec R-\vec{\sigma}_1
\vec{\sigma}_2 R^2)V_3} {12\mu_1\mu_2 R^2}. \label{4} \ee where
SDI potentials $V_i$ are expressed via field correlators as in
\cite{40}, see also definition of correlators in ({35})

\be
\frac{1}{R}\frac{dV_1}{dR}=-\int^{\infty}_{-\infty}d\nu\int^R_0
\frac{d\lambda}{R} \left (1-\frac{\lambda}{R}\right
)D^H(\lambda,\nu) \label{8} \ee \be
\frac{1}{R}\frac{dV_2}{dR}=\int^{\infty}_{-\infty}d\nu\int^R_0
\frac{\lambda d\lambda}{R^2} \left
[D^H(\lambda,\nu)+D_1^H(\lambda,\nu)+\lambda^2\frac{\partial
D_1^H}{\partial\lambda^2}\right ] \label{9} \ee \be
V_3=-\int^{\infty}_{-\infty} d\nu R^2\frac{\partial
D_1^H(R,\nu)}{\partial R^2} \label{10} \ee \be
V_4=\int^{\infty}_{-\infty}d\nu \left
(3D^H(R,\nu)+3D_1^H(R,\nu)+2R^2\frac{\partial D_1^H}{\partial
R^2}\right ) \label{11} \ee

 \be \frac{1}{R}\frac{d\varepsilon
(R)}{dR}=\int^{\infty}_{-\infty}d\nu\int^R_0 \frac{ d\lambda}{R}
\left
[D^E(\lambda,\nu)+D_1^E(\lambda,\nu)+(\lambda^2+\nu^2)\frac{\partial
D_1^E}{\partial\nu^2}\right ] \label{A3.34} \ee In the
deconfinement phase $D^E$ vanishes, while $D^H, D^E_1, D_1^H$ are
nonzero, and the main part of  $D^E_1, D_1^H$ is perturbative,
yielding the known from of SDI  essential at small distances. One
can assume (in agreement with lattice data \cite{13, 14}) that
$D_1^E, D_1^H$ do not change with temperature, which yields
$(D_{1pert}^E=D^H_{1 pert}=\frac{16\alpha_s}{3\pi x^4})$
$$\frac{1}{R} V'_{2pert}(R) = \frac{4\alpha_s}{3R^3},~~
\frac{1}{R} \varepsilon'_{pert}(R) = +\frac{4\alpha_s}{3R^3},~~
V'_{3pert}(R) = \frac{4\alpha_s}{R^3}$$ \be  V_{4pert}(R) =
\frac{32\pi \alpha_s}{3}\delta^{(3)}(\veR), ~~  V'_{1pert} =
0.\label{A3.35}\ee

Since NP colormagnetic fields at $T>T_c$ act only on states with
nonzero $L$, one can neglect $V_{4np}$. Writing $D_1^{E,H}
=D_{1pert}^{E,H} + D_{1np}^{E,H}$, one has from (\ref{A3.34}) (for
$T>T_c, \beta\equiv 1/T$ $$ \frac{1}{R} \varepsilon'_{np}
=\int^\beta_0 (1-y/\beta) dy D^E_{1np} (\sqrt{y^2+R^2}),$$
\be\frac{1}{R} V'_{2np} =\int^\beta_0 (1-y/\beta) dy D^H_{1np}
(\sqrt{y^2+R^2}) +2\int^\beta_0  d\nu (1-\nu/\beta) \int^R_0
\frac{\lambda d\lambda}{R^2} D^H(\lambda,\nu)\label{A3.36}\ee

Finally for $V'_1 = V'_{1np}$ one has from \cite{40} \be
\frac{1}{R} V'_1 = -2 \int^\beta_0 (1-\nu/\beta) d\nu \int^R_0
\frac{d\lambda}{R} \left(1-\frac{\lambda}{R}\right)
D^H(\sqrt{\lambda^2+\nu^2}).\label{A3.37}\ee

We have corrected  in (\ref{A3.35}),(\ref{A3.37}) equations from
\cite{40}, taking  into account the finiteness of the Euclidean
time  integration in the    temporal Green's functions at $T>0$.
As one can see, the presence of upper limit becomes essential only
when $\frac{1}{T} \la t_g$, where $t_g$ is the correlation  length
of the vacuum, $D^H(x) \sim e^{-|x|/t_g}$, for $|x|\gg t_g$, and
for $T=0$ one has $t_g \approx 1$ GeV$^{-1}$. Thus one expects,
that for $T<1$ GeV$\approx 6T_c$ the upper limit $\beta$ is
inessential and one can use the same integrals as in \cite{40},
i.e. replacing $\beta$ by $\infty$.

At large $R$ one has from (\ref{A3.36}), (\ref{A3.37}) \be
\frac{1}{R} V'_1(R)=-\sigma_s/R,~~ \frac{1}{R} V'_{2np} (R)
=\frac{\gamma_H}{R^2},~~R \to \infty \label{A3.38}\ee where
$\gamma_H$ is \be \gamma_H=\int^\infty_{-\infty} d\nu
\int^\infty_0 \lambda d\lambda D^H(\sqrt{\lambda^2+\nu^2})\approx
O(t_g\cdot \sigma_s).\label{A3.39} \ee Hence the main term at
large $R$ in $V_{SD} $({A3.29}) is \be V_{SD} (R\to\infty) \approx
-\left( \frac{\vesig_1 \veL_1}{2\mu_1\hat{ \mu}_1}
-\frac{\vesig_2\veL_2}{2\mu_2\hat{\mu}_2}\right)
\frac{\sigma_s}{R}.\label{A3.40}\ee

For equal mass particles $\mu_1=\mu_2=\mu,~~ \veL_2
=-\veL_1=-\veL, ~~ \veS =\frac12 (\vesig_1+\vesig_2)$ and one
obtains \be V_{SD} (r\to \infty) \approx
-\frac{\sigma_s\veS\veL}{2\mu\hat{\mu}R} \left(1+
O\left(\frac{t_g}{R}\right)\right) + O\left(
\frac{\alpha_s}{\mu^2R^3}\right),\label{A3.41}\ee where
$\hat{\mu}= \mu +2\int^1_0 d\beta (\beta-\frac12)^2 \nu(\beta)
\approx  \mu+\frac{\sigma R}{6}$.

It is clear that interaction (\ref{A3.41}) can produce infinite
number of bound states for $\veS\veL>0$, i.e. for $S=1$ and
$J=L+1$, independently of the dynamical mass $\mu$. In this way
(\ref{A3.41}) embodies the suggestion made in \cite{19}.

One can also estimate the interaction (\ref{A3.41}) for the free
quark with the momentum $p$, when $\mu\approx p$ and $\veL =[\ver
\times \vep] = [\verho\times \vep]$, where $\rho$ is the impact
parameter to the neighboring antiquark.

One has $|V_{SD} | \approx \frac{\sigma_s}{p^2} \frac{\rho\cdot
p}{R}\sim \frac{\sigma_s}{p}$, and the standard plasma interaction
ratio $\Gamma$ for average momentum $\lan p \ran \sim T$ \be
\Gamma_q =\frac{|V_{SD}|}{T_{kin}} \approx
\frac{\sigma_s}{p^2}\approx  \frac{\sigma_s}{T^2}; ~~ \Gamma_g
=2\cdot \frac94 \frac{\sigma_s}{T^2}\label{A3.42}\ee where in
$\Gamma_g$ the first factor (2) is due to gluon spin  which is
twice the quark spin, and the second is Casimir scaling factor.

One can see that  for $T_c\leq T\leq 2T_c$ this ratio is larger
than unity, implying strong interaction  due to magnetic vacuum,
especially among gluons. Therefore gluons (and quarks) cannot be
considered as a gas of weakly interacting objects, e.g. for
$T\approx T_c \approx 0.2$ GeV one has $\Gamma_q \approx 5,~~
\Gamma_g\approx 20$ and this is closer to a liquid rather than a
gas.

\end{document}